\documentstyle[prd,aps]{revtex}

\def\now{\ifcase\month\or Jan.\or Feb.\or March\or April\or May\or
June\or July\or Aug.\or Sept.\or Oct.\or Nov.\or Dec.\fi
\space\number\day,
\number\year\space (\number\time)}

\newcommand{\vs}{v_{\rm s}}
\newcommand{\va}{\tilde{B}_0}

\newcommand{\mn}{^{\mu\nu}}
\newcommand{\p}{\partial}
\def\o{\over}
\newcommand{\aoa}{\frac{\dot{a}}{a}}
\def\simle{\mathrel{{}^<_\sim}}

\begin{document}
\title{Damping of Cosmic Magnetic Fields}
%
\author{Karsten Jedamzik}
\address{Max-Planck-Institut fuer Astrophysik}
\address{85748 Garching bei Muenchen, Germany}
\author{Vi\v{s}nja Katalini\'c and Angela V. Olinto}
\address{Department of Astronomy and Astrophysics and Enrico 
Fermi Institute}
\address{University of Chicago, 5640 S. Ellis Ave, Chicago, IL 60637}
%
\maketitle

\begin{abstract}
We examine the evolution of magnetic fields in an expanding fluid
composed of matter and radiation with particular interest in the
evolution of cosmic magnetic fields. We derive the propagation
velocities and damping rates for relativistic and non-relativistic
fast and slow magnetosonic, and Alfv\'en waves in the presence of
viscous and heat conducting processes. The analysis covers all MHD
modes in the radiation diffusion and the free-streaming regimes. When
our results are applied to the evolution of magnetic fields in the
early universe, we find that cosmic magnetic fields are damped from
prior to the epoch of neutrino decoupling up to recombination.
Similar to the case of sound waves propagating in a demagnetized
plasma, fast magnetosonic waves are damped by radiation diffusion on
all scales smaller than the radiation diffusion length. The
characteristic damping scales are the horizon scale at neutrino
decoupling ($M_{\nu} \approx 10^{-4} M_{\odot}$ in baryons) and the
Silk mass at recombination ($M_{\gamma} \approx 10^{13} M_{\odot}$ in
baryons).  In contrast, the oscillations of slow magnetosonic and
Alfv\'en waves get overdamped in the radiation diffusion regime,
resulting in frozen-in magnetic field perturbations. Further damping
of these perturbations is possible only if before recombination the
wave enters a regime in which radiation free-streams on the scale of
the perturbation. The maximum damping scale of slow magnetosonic and
Alfv\'en modes is always smaller than or equal to the damping scale of
fast magnetosonic waves, and depends on the magnetic field strength
and its direction relative to the wave vector.

Our findings have multifold implications for cosmology. The
dissipation of magnetic field energy into heat during the epoch of
neutrino decoupling ensures that most magnetic field configurations
generated in the very early universe satisfy big bang nucleosynthesis
constraints. Further dissipation before recombination constrains
models in which primordial magnetic fields give rise to galactic
magnetic fields or density perturbations. Finally, the survival of
Alfv\'en and slow magnetosonic modes on scales well below the Silk
mass may be of significance for the formation of structure on small
scales.


\end{abstract}

\section{Introduction}
 
In an attempt to explain the origin of galactic magnetic fields
through the amplification of primordial fields, several authors have
considered scenarios for generating magnetic fields in the early
universe \cite{fieldcreation}. In such scenarios, one attempts to
generate fields which will be sufficiently large after recombination
at least to seed galactic dynamos and at best to produce galactic
fields without dynamo amplification. It is generally assumed that
after a primordial field is generated in the early universe, it
becomes frozen into the cosmic plasma and redshifts by flux
conservation with the expansion of the universe ($B \propto a^{-2}$;
$a(t)$ is the cosmic scale factor). This assumption is usually
justified by noting that the cosmological plasma is highly conductive
and magnetic diffusion is insignificant.

In this paper, we show that this simple picture of magnetic field
evolution is incorrect: at certain epochs in the early universe,
particularly during recombination and neutrino decoupling, magnetic
field energy is converted into heat through the damping of
magneto-hydrodynamic (MHD) modes. The damping is caused by dissipation
in the fluid, which arises from the finite mean free path of photons
or neutrinos.

The physical process by which the MHD modes are damped is analogous to
that involved in the damping of density fluctuations around
recombination \cite{S}, and around neutrino decoupling
\cite{M68}. Studies of the damping of density fluctuations with no
magnetic fields present show that, in the diffusive regime (when the
scales of interest are much larger than the mean free path of photons
or neutrinos, $\l_{\rm mfp}$), the effective viscosity and heat
conductivity arising from the finite mean free path cause the damping
of acoustically oscillating density perturbations. Since with the
expansion of the universe the mean free path of the decoupling
particles grows faster than the wavelength of an oscillatory mode, all
modes whose wavelengths are smaller than the mean free path around
decoupling have previously been in the diffusion regime. The rate of
damping in this regime ensures that a wave is significantly damped
before the mean free path of the decoupling particles becomes
comparable to the wavelength of the mode. For this reason, the
investigation of damping in the diffusion regime yields a reasonable
estimate of the final damping scales of density fluctuations.

However, when magnetic fields are added to the fluid, the existence of
different MHD modes---Alfv\'en, fast magnetosonic, and slow
magnetosonic waves---adds complexity to the problem. We show that
while fast magnetosonic waves (which include sound waves) damp
efficiently in the diffusion regime by the described process, slow
magnetosonic and Alfv\'en waves may survive damping by diffusion. Slow
magnetosonic and Alfv\'en modes oscillate with frequencies which
depend on the strength of the background magnetic field and on its
direction relative to the mode's wave vector, and are in general
different from the frequency of sound waves of the same wavelength. In
the case of a weak background magnetic field or a large angle between
the background field and the wave vector, the frequency can be slow
enough for the damping by viscosity to overcome the oscillation,
producing behavior which resembles an overdamped oscillator and causes
the actual damping of the amplitude to be inefficient. The overdamped
slow magnetosonic and Alfv\'en modes therefore survive diffusion
damping. However, they undergo additional damping if, with the
expansion of the universe, they enter the so called free-streaming
regime, i.e. the mean-free-path grows to be much larger that the
wavelength of a mode. As a consequence, whereas fast magnetosonic
modes are damped mostly when radiation is diffusing, slow magnetosonic
and Alfv\'en modes are also significantly damped when radiation is
free-streaming. Therefore, when studying the damping of all MHD modes
in order to estimate their damping scales, it is necessary to
investigate both the free-streaming and the diffusion regimes even
before the final stages of the decoupling process.

The damping of MHD modes which causes the dissipation of magnetic
energy can be illustrated with the following picture: as long as there
exist spatially tangled magnetic fields, Lorentz forces accelerate the
fluid, setting up oscillations about a force-free field configuration;
the induced motions are damped by the effective viscosity of photons
or neutrinos; this causes the exponential decrease in the amplitude of
the oscillations and thus results in the straightening of magnetic
field lines towards a force-free configuration. After the cosmological
magnetic fields undergo this damping process they have little
structure on scales below a characteristic damping scale, and the
magnetic energy density in such primordial fields is much smaller than
that expected from the simple redshift argument above.

In this paper we follow the evolution of MHD modes and derive their
propagation velocities and damping rates both in the diffusion and
free-streaming regimes during the decoupling of photons and
neutrinos. The existence of highly relativistic particles with mean
free path much shorter than the wavelength of a MHD mode (e.g.,
photons and leptons) requires the use of relativistic MHD. In the
radiation diffusion regime, studied in \S 2, we develop a relativistic
description of viscous expanding fluids with magnetic fields, while in
the free-streaming case, \S 3, the effects of the photons or neutrinos
are included through heat exchange and a drag force which they exert
on the fluid. This procedure allows us to calculate, in \S 4, the
maximum damping lengths after the epochs of neutrino decoupling and
recombination. Our results may be applied to other astrophysical
environments where MHD waves propagate in a viscous fluid, since in
the derivation of the dispersion relations we leave the sources of
viscosity and heat conductivity unspecified.

\section{Damping of Magnetohydrodynamic Modes in the Radiation
Diffusion Regime}

When the mean free paths of all interacting particle species are
shorter than the wavelength of the MHD mode we are interested in
($\lambda \gg l_{\rm mfp}$), it is adequate to study the evolution of
a single fluid and account for the effect of the diffusing particles
by introducing shear viscosity, bulk viscosity, and heat conductivity
into the fluid equations \cite{W71}. 
In order to calculate the damping of MHD modes following this
approach, we derive linearized relativistic MHD equations of an
expanding dissipative fluid. We start by reviewing the equations for a
non-ideal relativistic fluid in \S 2.1, and add the electromagnetic
contributions to the fluid equations in \S 2.2. In \S 2.3 we calculate
the propagation velocities and damping rates for all MHD modes. Our
results are applicable for general viscous relativistic and
non-relativistic plasmas, as long as the pressure is dominated by
radiation pressure.

Throughout the paper we assume that the magnetic field can be
decomposed into a large magnitude background component ${\bf B}_0({\bf
x},t)$, and a small perturbation, ${\bf b}({\bf x},t)$. We
additionally assume that the curl of the background component is
negligible when compared to the curl of the perturbations. These two
assumptions allow us to solve for the damping of MHD modes
analytically. 

The use of scalar viscosities and heat conductivity implicitly
neglects any anisotropies in these quantities due to the presence of
the magnetic field. Further, since our equations are derived for an
isotropically, homogeneously, and adiabatically expanding plasma, the
background magnetic field, ${\bf B}_0$, is required to have vanishing
spatial average on sufficiently large scales, $\langle {\bf
B}_0\rangle =0$. In our derivation we also neglect gravitational
forces because the scales of interest are smaller than the Jeans mass
scale, and we assume the plasma to be infinitely conducting which is
an excellent approximation for most astrophysical plasmas and for the
early universe (see, e.g., Ref.~\cite{CO}).

\subsection{Relativistic Imperfect Fluids}

We consider the evolution of a non-ideal, relativistic fluid in a
homogeneously and isotropically expanding background using the
spatially flat Robertson-Walker metric $g_{\mu\nu}={\rm diag}(1, -a^2,
-a^2, -a^2)$ and comoving coordinates $x^{\mu}$. The time dependent
scale factor $a(t)$ provides the connection between proper (physical)
coordinates ${x'}^\mu$ and the comoving coordinates: ${x'}^0=x^0$ and
${x'}^i=ax^i$ (Greek indices run from 0 to 3 whereas Latin indices run
from 1 to 3).

The relativistic fluid is described by the energy-momentum tensor
\begin{equation} T\mn = T\mn_{ I} + \tau^{\mu\nu} +
                        T\mn_{{\rm EM}},  \label{eq:tensor}
\end{equation}
which is separated into three parts: the ideal fluid tensor
$T\mn_{I}$, the non-ideal fluid part $\tau^{\mu\nu}$, which accounts
for dissipation, and the electromagnetic energy-momentum tensor
$T\mn_{{\rm EM}}$ (added in \S 2.2).  The equations of fluid dynamics can
be derived from energy-momentum conservation
\begin{equation}
  T\mn{}_{;\nu} = 0.   \label{eq:conserv_Tmn}
\end{equation}
In comoving coordinates Eq.~\ref{eq:conserv_Tmn} becomes
\begin{equation}
  \frac{\p T^{0\mu}}{\p x^{\mu}} +
  \dot{a}a\biggl(\sum_{i=1,3}T^{ii}\biggr) +
  3\biggl(\aoa\biggr)T^{00} = 0,    \label{eq:tmn_energy}
\end{equation}
and
\begin{equation}
  \frac{\p T^{i\mu}}{\p x^{\mu}} +
  5\biggl(\aoa\biggr) T^{i0} = 0,  \label{eq:tmn_momentum}
\end{equation}
with the dot representing a derivative with respect to time
$x^0$. 

The energy-momentum tensor for an ideal fluid is
\begin{equation}
  T\mn_{I} = (\rho+p) U^{\mu} U^{\nu} - p g\mn,
\end{equation}
where $\rho$, $p$, and $U^{\mu}$ are the total energy density,
the total pressure, and the four velocity of the fluid, respectively.
The non-ideal contributions to the fluid energy-momentum tensor can
be written as \cite{W72}
\begin{eqnarray}
  \tau\mn &=& \eta\Bigl(U^{\mu ;\nu}
            + U^{\nu;\mu}
            - U^{\mu} U^{\lambda} U^{\nu}{}_{;\lambda}
            - U^{\nu} U^{\lambda} U^{\mu}{}_{;\lambda}\Bigr) 
\nonumber \\
&&{} + \Bigl(\xi - \frac{2}{3} \eta\bigr) U^{\lambda}{}_{;\lambda}
              \bigl(g\mn - U^{\mu} U^{\nu} \Bigr) \nonumber \\
       &&{} + \kappa \biggl[U^{\mu} \biggl( \frac{\p T}{\p x_{\nu}}
            - T U^{\nu}{}_{;\lambda} U^{\lambda}\biggr)
            + U^{\nu}\biggl( \frac{\p T}{\p x_{\mu}}
        - T U^{\mu}{}_{;\lambda} U^{\lambda}\biggr)
        - 2 U^{\mu} U^{\nu} \frac{\p T}{\p x_{\lambda}} U_{\lambda}
                \biggr]. \label{eq:tau}
\end{eqnarray}
In this expression, $T$ stands for temperature and $\eta$, $\xi$, and
$\kappa$ are shear viscosity, bulk viscosity, and heat conductivity
respectively.  

The effective viscosities and heat conductivity for either photons
or neutrinos are given by \cite{MS65,M68,W71}:
\begin{equation}  
\eta={4\over 15}g{\pi^2\over 30}T^4 l_{\rm mfp}\,  , 
			\label{eq:shear} 
\end{equation} 

\begin{equation} 
\xi=4g{\pi^2\over 30}T^4\biggl[{1\over 3}-\biggl({\partial
p\over\partial\rho}\biggr)_n\biggr] l_{\rm mfp}\, , 
\end{equation}

\begin{equation} 
\kappa={4\over 3}g{\pi^2\over 30}T^3 l_{\rm mfp}\, ,
		\label{eq:heat} 
\end{equation} 
where $n$ is the number density of the conserved particles in the
fluid and $g$ is the statistical weight of the diffusing particles. 

The exact form of $\tau^{\mu\nu}$ is partially a matter of definition
since, in relativistic fluid mechanics, the fluid velocity can be
defined either by the flow of conserved particles \cite{W72} or by
the flow of energy \cite{LL75}. These definitions coincide in
non-relativistic fluid mechanics where the rest mass of particles
dominates the total energy. In our case, a relativistic one-fluid
approximation, the charged and strongly interacting particles
(protons, neutrons, electrons, etc.)  which compose the fluid are all
perfectly coupled and have the same velocity as the conserved particle
number, the baryon number. The energy flow may differ from the
particle flow, however, due to the energy transported by the
imperfectly coupled neutrinos and photons.  We choose to follow the
velocity of the charged particles (and, therefore, the flow of baryon
number) which appears explicitly in the magneto-hydrodynamic equations
below.

The conservation of particle number can be written as
\begin{equation}
n^{\mu}_{\,\,\,\, ;\mu}=0 \ , \label{eq:nmu_conserve}
\end{equation}
where $n^{\mu} =n U^{\mu}$ is the particle number four current with
$n$ the proper number density of particles.  The particle number we
follow is the net baryon number, $n^b$, which is conserved for
temperatures below the electroweak transition.

We can now derive the linearized equations of ordinary
relativistic fluid dynamics in an expanding universe from
Eq.~\ref{eq:conserv_Tmn} -~\ref{eq:nmu_conserve}, by expanding the
fluid variables around their background values
\begin{eqnarray}
\rho({\bf x},t) &=& \rho_0(t)+\rho_1({\bf x},t), \\
   p({\bf x},t) &=& p_0(t)+p_1({\bf x},t),\\
   T({\bf x},t) &=& T_0(t)+T_1({\bf x},t),\\
 n^b({\bf x},t) &=& n^b_0(t)+n^b_1({\bf x},t),\\
        U^{\mu} &=& U^{\mu}_0+U^{\mu}_1 \ .
\end{eqnarray}
The four velocity is that of a stationary
fluid element (with respect to the comoving frame) plus a small 
velocity perturbation
\begin{equation}
\quad\ U^{\mu}_0=\bigl(1,0,0,0\bigr),\quad
U^{\mu}_1=\bigl(0,{{\bf v}\over a}\bigr). \label{eq:u}
\end{equation}
We choose $U_1^{\mu}$ in this particular form so that the fluid
velocity in proper coordinates, ${\bf v}'=(\dot{a}/a){\bf x}'+{\bf
v}$, corresponds to an isotropic expansion plus an additional peculiar
velocity ${\bf v}$. We consider fluids in which the peculiar
velocities are much smaller than the speed of light, e.g. $|{\bf
v}|\ll 1$, where velocities are measured in units of the speed of
light. Although the fluid velocities are small, a relativistic
treatment is necessary to adequately account for the presence of
relativistic particles (e.g., photons and neutrinos).

Evaluating
equations~\ref{eq:conserv_Tmn}~and~\ref{eq:nmu_conserve} to
lowest order in the fluid variables, we obtain
\begin{equation}
   \frac{\p \rho_0}{\p t} + 3\biggl(\aoa \biggr)\bigl(\rho_0+p_0\bigr)
   = 9\xi \biggl(\aoa \biggr)^2 ,  \label{eq:entropy0}
\end{equation}
which represents conservation of entropy when $\xi =0$, and
\begin{equation}
\frac{\p n_0^b}{\p t} + 3\biggl({\aoa}\biggr)n_0^b = 0
, \label{eq:number0}
\end{equation}
represents conservation of baryon number.  If we expand the energy
momentum tensor and baryon number four-current to first order in the
perturbation variables $U_1^{\mu}$, $\rho_1$, $p_1$, $T_1$, and
$n_1^b$, use Eq.~\ref{eq:tmn_energy}--\ref{eq:tmn_momentum}, and
subtract the zeroth order solution, we obtain
\begin{equation}
\frac{\p\rho_1}{\p t}
     + \bigl(\rho_0+p_0\bigr) \frac{1}{a}{\bf\nabla\cdot v}
     + 3\biggl({\aoa}\biggr)\bigl(\rho_1+p_1-2\xi{\bf\nabla\cdot
v}\bigr)
     - \frac{\kappa}{a^2} \nabla^2 T_1
     - \frac{\kappa}{a^2} \frac{\p}{\p t}\bigl(a T_0 {\bf\nabla\cdot
v}\bigr)
     = 0,    \label{eq:general_energy} \end{equation}
\begin{equation}
\frac{1}{a^4}\frac{\p}{\p t}\Biggl[ a^4\bigl(\rho_0+p_0\bigr)
{\bf v}
  -\kappa a^3\biggl[\frac{\p}{\p t}\bigl(a T_0{\bf v}\bigr)
  + {\bf\nabla} T_1\biggr]
  - 3\xi a^3\dot{a}{\bf v}\Biggr] + \frac{1}{a}{\bf\nabla} p_1  
  - \frac{\eta}{a^2} \nabla^2 {\bf v}
  -\frac{1}{a^2}\bigl(\xi+\frac{1}{3}\eta\bigr){\bf\nabla}
({\bf\nabla\cdot v})
  = 0,    \label{eq:general_momentum}
\end{equation}
and
\begin{equation}
\frac{\p n_1^b}{\p t} + \frac{n^b_0}{a}{\bf \nabla\cdot v}
   + 3\biggl(\aoa \biggr)n_1^b= 0 .   \label{eq:number1}
\end{equation}

These equations form a complete set describing the evolution of
a non-ideal fluid; Eq.~\ref{eq:general_energy} represents the
first law of thermodynamics in local form,
Eq.~\ref{eq:general_momentum} is the relativistic version of
Euler's equation, and Eq.~\ref{eq:number1} represents the
conservation of baryon number.

\subsection{Magneto-Hydrodynamics with Dissipation}

We now include the electromagnetic fields. In an inertial frame 
(denoted by an overhat) the Maxwell tensor  has the form
\begin{equation}
  \hat{F}^{\mu\nu} = \left( \begin{array}{cccc}
         0 & E_x & E_y  & E_z  \\
         -E_x & 0 & B_z & -B_y \\
         -E_y & -B_z & 0 & B_x \\
         -E_z & B_y & -B_x & 0
         \end{array} \right),
\end{equation}
where $E_i$ and $B_i$ are the electric and magnetic fields as 
seen by an
observer in the inertial frame. 
The Maxwell tensor
in comoving coordinates ($x^{\mu}$) can be derived 
from the Maxwell tensor in inertial coordinates
($\hat{x}^{\mu}$) by using the transformation rules for
tensors
\begin{equation}
F^{\mu\nu}=\Lambda^{\mu}{}_{\lambda}\Lambda^{\nu}{}_{\sigma}\hat{F}^{\lambda\sigma}\ ,
\label{eq:transform}
\end{equation}
where
\begin{equation}
\Lambda^{\mu}{}_{\nu}={\p x^{\mu}\o\p \hat{x}^{\nu}}\ .
\end{equation}

The coordinate transformation which transforms the locally Minkowski metric
$\hat{g}^{\mu\nu}={\rm diag}(1,-1,-1,-1)$ into the Robertson-Walker metric
$g^{\mu\nu}={\rm diag}(1, -1/a^2, -1/a^2, -1/a^2)$ has
\begin{equation}
\Lambda^{\mu}{}_{\nu}={\rm diag}(1\, ,\, 1/a\, ,\, 1/a\, ,\, 1/a\, )\ . 
\label{eq:lambda}
\end{equation} 
Thus, in the comoving basis $F\mn$ is
\begin{equation}
F\mn = \left(\begin{array}{cccc}
     0 & E_x/a & E_y/a & E_z/a  \\
         -E_x/a & 0 & B_z/a^2 & -B_y/a^2 \\
         -E_y/a & -B_z/a^2 & 0 & B_x/a^2 \\
         -E_z/a & B_y/a^2 & -B_x/a^2 & 0
         \end{array} \right).     \label{eq:fmn}
\end{equation}
The equations of motion for 
the electromagnetic
fields are Maxwell's equations
\begin{equation}
F\mn{}_{;\nu} = 4\pi J^{\mu}
\end{equation}
and
\begin{equation}
{\p\o\p x^{\lambda}}F_{\mu\nu}+{\p\o\p x^{\nu}}F_{\lambda\mu}+{\p\o\p
x^{\mu}}F_{\nu\lambda}=0\ ,\label{eq:Maxwell}
\end{equation}
where $J^{\mu}$ is the electric four current.

In the limit of infinite
electrical conductivity, the electric field in the rest frame 
of the charged particles vanishes
\begin{equation}
  E^\mu = F\mn U_\nu = 0.   \label{eq:mhd}
\end{equation}
This condition, evaluated in the comoving frame using
equations~\ref{eq:u} and~\ref{eq:fmn}, reads
\begin{equation}
  {\bf E} = - {\bf v}\times{\bf B}. \label{eq:mhd1}
\end{equation}

We decompose the magnetic field 
into its background value, ${\bf
B}_0$, and a small-amplitude perturbation, ${\bf b}({\bf x},t)$
\begin{equation}
{\bf B}({\bf x},t) = {\bf B}_0({\bf x}, t) + {\bf b}({\bf x},t), 
\end{equation}
and   impose
the following conditions 
\begin{equation}
{\bf b}({\bf x},t) \ll {\bf B}_0({\bf x},t)\ ,
\end{equation}
\begin{equation}
\nabla\times{\bf B}_0({\bf x},t) \ll \nabla\times{\bf b}({\bf x},t)\ .
\end{equation}
to linearize our equations.
We can now derive the relevant Maxwell's equations to zeroth and first
order in the small quantities ${\bf v}$ and ${\bf b}$ by using
Eq.~\ref{eq:fmn},~\ref{eq:Maxwell},~and~\ref{eq:mhd1}. This yields
\begin{eqnarray}
\nabla\cdot{\bf b} & = & 0, \label{eq:maxw1} \\
\frac{1}{a^2} \frac{\p}{\p t} (a^2{\bf b})
   & = & \frac{1}{a} \nabla\times ({\bf v}\times {\bf B}_0),
                                          \label{eq:maxw2} \\
\frac{\p}{\p t}(a^2 {\bf B}_0) & = & 0.   \label{eq:maxw3}
\end{eqnarray}
Equation~\ref{eq:maxw3} shows that the background field
${\bf B}_0$, by flux conservation, redshifts as~$1/a^2$ with the expansion
of the fluid.

To complete the system of equations needed to describe the evolution
of the fluid
in the presence of electromagnetic fields, we must add
the contribution from the
electromagnetic energy-momentum tensor $T\mn_{\rm EM}$
to the conservation of energy-momentum Eqs.~\ref{eq:tmn_energy}
and \ref{eq:tmn_momentum}.
The energy-momentum
tensor for electromagnetism is
\begin{equation}
T\mn_{\rm EM} = \frac{1}{4\pi}\Bigl\{ F^{\mu\sigma}
F^{\nu}{}_{\sigma} -
       \frac{1}{4} g\mn F^{\sigma\rho}F_{\sigma\rho}\Bigr\},
\end{equation}
which in comoving coordinates becomes
\begin{equation}
T\mn_{\rm EM}= \frac{1}{4\pi}\left(
\begin{array}{cc}
A &  {\bf S}  \\
 {\bf S}  & \sigma^{ij}
\end{array}\right),
\end{equation}
with
\begin{equation}
A = \frac{\bigl({\bf E}^2+{\bf B}^2\bigr)}{2},
    \quad {\bf S}=\frac{\bigl({\bf E\times B}\bigr)}{a}, \ \ {\rm
and }
   \ \ \sigma^{ij}=\frac{1}{a^2}\Bigl(-E_iE_j-B_iB_j
       +\frac{1}{2}\delta_{ij}\bigl({\bf E}^2+{\bf B}^2\bigr)
\Bigr).
\end{equation}

We can now evaluate the contributions from the electromagnetic stresses
to the conservation of entropy Eq.~\ref{eq:entropy0}, 
the first law of thermodynamics
in local form Eq.~\ref{eq:general_energy}, 
and the relativistic version of Euler's equation 
Eq.~\ref{eq:general_momentum}.
The contribution to the left-hand-side of the zeroth order
Eq.~\ref{eq:entropy0} is
\begin{equation}
\frac{1}{8\pi a^4}\frac{\p}{\p t}(a^4 {\bf B}_0^2),\label{eq:mag_entropy0}
\end{equation}
while to first order in ${\bf v}$ and ${\bf b}$,
the electromagnetic stresses to be added to the left-hand-sides of
Eqs.~\ref{eq:general_energy}~and~\ref{eq:general_momentum} are
\begin{equation}
\frac{1}{4\pi}\biggl[\frac{1}{a^4} \frac{\p}{\p t}
\bigl(a^4{\bf b}\cdot
               {\bf B}_0\bigr)
 + \frac{{\bf B}_0^2}{a} \bigl({\bf\nabla\cdot v}\bigr)
 - \frac{1}{a} \bigl({\bf B}_0\cdot {\bf \nabla}\bigr)
\bigl({\bf v\cdot
              B}_0\bigr)\biggr]\ .             \label{eq:mag_energy}
\end{equation}
for Eq.~\ref{eq:general_energy}, and
\begin{equation}
\frac{1}{4\pi a^4} \frac{\p}{\p t} \biggl(a^4{\bf B}_0\times\bigl(
{\bf v}           \times {\bf B}_0\bigr)\biggr)
  + \frac{1}{4\pi a} \biggl[{\bf B}_0{\bf\times}\bigl(
{\bf\nabla\times b}
          \bigr)\biggr]\ ,            \label{eq:mag_momentum}
\end{equation} 
for Eq.~\ref{eq:general_momentum}.  After substituting Maxwell's
equations (Eq.~\ref{eq:maxw1}--\ref{eq:maxw3}) into
Eq.~\ref{eq:mag_energy}, we find that the electromagnetic contribution
to Eq.~\ref{eq:general_energy} is zero.  Similarly,
Eq.~\ref{eq:mag_entropy0} is identical to zero by virtue of
Eq.~\ref{eq:maxw3} so that both Eq.~\ref{eq:entropy0}
and~\ref{eq:general_energy} are unmodified.  The only coupling between
the field and the fluid to first order occurs through the velocity of
charged particles and the curl of the magnetic field.

Note that the first term of Eq.~\ref{eq:mag_momentum} is only
important in the relativistic limit. For a mode with frequency
$\omega$ and wavenumber $k$, Eq.~\ref{eq:maxw2} implies $\omega b\sim
k {\bar v} B$.  Therefore, the relative contribution of the first term
in Eq.~\ref{eq:mag_momentum} compared to the second term is of order
$(\omega /k)^2$. Hence, the first term can only be neglected when the
group velocity of a mode, $\p\omega / \p k\approx \omega / k$, is much
smaller than the speed of light.

To first order in the quantities $T_1$, $n_1^b$, $\rho_1$, $p_1$, $\bf
v$, $\bf b$, equations~\ref{eq:general_energy}, \ref{eq:number1},
\ref{eq:maxw1}--\ref{eq:maxw3}, together with the equation obtained by
adding Eq.~\ref{eq:mag_momentum} to the left-hand-side of
Eq.~\ref{eq:general_momentum} describe magneto-hydrodynamics in an
expanding fluid.  The following definitions help to rewrite our
equations into a more convenient form
\begin{equation}
\delta \equiv {T_1\o T_0}\ ,\quad \Delta\equiv {n_1^b\o n_0^b}\ , \label{eq:7}
\end{equation}

\begin{equation}
\tilde{\bf b}\equiv {{\bf b}\o \bigl(4\pi (\rho_r+p_r) 
\bigr)^{1\o 2}}\ ,\quad
\tilde{\bf B}_0\equiv {{\bf B}_0\o \bigl(4\pi (\rho_r+p_r) 
\bigr)^{1\o 2}}\ ,
\label{eq:8}
\end{equation}

\begin{equation}
\eta^{\prime}\equiv {\eta\o 3  (\rho_r+p_r)}\ ,\quad
\xi^{\prime}\equiv {\xi\o 3 (\rho_r+p_r)}\ ,\quad
\kappa^{\prime}\equiv {\kappa T\o 3 (\rho_r+p_r)}\ , \label{eq:9}
\end{equation}
where $\rho_r=g \pi^2 T_0^4/30 $ and $p_r=\rho_r/3$ are the
average energy density and pressure of relativistic particles,
and  $g$ is the total statistical weight of relativistic
particles.

For a fluid comprised of baryons and relativistic particles
(e.g., photons,
neutrinos, $e^{\pm}$-pairs, etc.),
 the energy density and pressure up to first order
in the small quantities are
given by
\begin{eqnarray}
  \rho &=& \rho_0 + \rho_1                                
\nonumber\\
       &=& \rho_r\bigl(1+4\delta\bigr) + \rho^b\bigl(1+\Delta\bigr) \, , \\
  p &=& p_0 + p_1                                         
\nonumber\\
    &=& p_r\bigl(1+4\delta\bigr)\ ,\label{eq:pressure}
\end{eqnarray}
where $\rho_b = n^b_0 m_{\rm N}$ is the baryon energy density, and
$m_{\rm N}$ is the nucleon rest mass. In writing
Eq.~\ref{eq:pressure}, we assume that baryonic pressure is negligible
in comparison to radiation pressure.  In this case, and when $\xi
=0$, Eq.~\ref{eq:entropy0} and Eq.~\ref{eq:number0} imply simple
redshift relations for the temperature $T_0\sim 1/a$ and the baryon
number density $n_0^b\sim 1/a^3$.  We also define
\begin{equation} 
R (t) \equiv {3\rho_b\o 4\rho_r}  \label{eq:15} 
\end{equation}
as a measure of the relative importance of baryon mass density with
respect to energy density in relativistic particles.  For $R
\rightarrow 0$, both energy density and pressure are dominated by
relativistic particles, whereas for $R \gg 1$ the energy density is
dominated by the baryon rest mass and the pressure is dominated by
radiation.

In terms of the newly defined variables, the equations of
magneto-hydrodynamics become
\begin{equation}
\dot{\delta}+ 36 \xi^{\prime} \biggr(\aoa\biggr)^2 \delta + {1\o 3
a}{\bf\nabla\cdot v}
-6\biggl({\aoa}\biggr) {\xi^{\prime}\o
a}{\bf\nabla\cdot v}-{\kappa^{\prime}\o a^2}\nabla^2\delta
-{\kappa^{\prime}\o
a}{\p\o\p t}{\bf\nabla\cdot v}=0\ , \label{eq:finalfluid0}
\end{equation}
\begin{eqnarray}
\biggl({\p\o\p t}+36\xi'\biggr(\aoa\biggl)^2\biggr) 
\biggl((1+R){\bf v} 
 - 3\kappa^{\prime}\bigl(\dot{\bf v}+{1\o
a}{\bf\nabla}\delta\bigr)
&-& 9\xi^{\prime}\Bigl({\aoa}\Bigr){\bf v}\biggr)
+ {1\o a}{\bf\nabla}\delta 
-{3\eta^{\prime}\o a^2}
\biggl(\nabla^2
{\bf v}+{1\o 3}{\bf\nabla}\bigl({\bf\nabla\cdot v}\bigr)\biggr)
\nonumber \\
& &{} -{3\xi^{\prime}\o
a^2}{\bf\nabla}\bigl({\bf\nabla\cdot v}\bigr)
+ \tilde{\bf B}_0\times \biggl({\p {\bf v}\o\p
t}\times\tilde{\bf B}_0\biggr)
+{1\o a}\tilde{\bf
B}_0{\bf\times}\bigl({\bf\nabla\times\tilde{b}}\bigr)=0\ ,
                            \label{eq:finalfluid1}
\end{eqnarray}

\begin{equation}
{\p\o\p t}\tilde{\bf b}={1\o a}{\bf\nabla\times}\bigl({\bf
v\times\tilde{B}}_0\bigr)\ , \label{eq:finalfluid2}
\end{equation}

\begin{equation}
{\bf\nabla\cdot\tilde b}=0\ , \label{eq:finalfluid3}
\end{equation}

\begin{equation}
{\p\o\p t}\tilde{\bf B}_0=0\ . \label{eq:finalfluid4}
\end{equation}

\subsection{Dispersion Relations}

In order to calculate propagation velocities and damping rates, we
first derive the dispersion relations for the different MHD modes by
Fourier transforming all perturbative variables ($\delta$, $\Delta$,
$\bf b$, and $\bf v$, generically represented by $\Phi$ below) using
the convention
\begin{equation} 
\Phi ({\bf x}, t) = \int\!d^3{\bf k}\, \Phi ({\bf k}, t)
      \exp \bigl(i{\bf k}\cdot{\bf x}\bigr) \label{eq:F_transf}
\end{equation} 
in which ${\bf k}$ is a constant comoving wave vector. 

The time dependence of $\Phi ({\bf k}, t)$ is modified by the
expansion of the fluid which introduces a time variation into the
frequency and the amplitude of the modes. With this in mind it is
convenient to write
\begin{equation} 
	\Phi ({\bf k}, t) = \Phi_{\bf k}(t) \exp \Bigl[\int i\omega(t)\, 
				dt \Bigr]. \label{eq:time_depend}
\end{equation}
The decrease of the amplitude due to damping is included in the
exponential part through imaginary solutions for $\omega$ while the
explicit time dependence of $\Phi_{\bf k}(t)$ acounts only for the
effects of the expansion.

The system of equations resulting from the substitution of
Eq.~\ref{eq:F_transf} into
Eq.~\ref{eq:finalfluid0}--\ref{eq:finalfluid4} is solved separately
for the different MHD modes: Alfv\'en waves, for which the density and
the temperature of the fluid are uniform and the velocity of the fluid
is perpendicular to the background magnetic field (${\bf k}=k\hat{\bf
x}, {\bf B}_0= B_x\hat{\bf x}+B_z\hat{\bf z}, {\bf b}=b\hat{\bf y},
{\bf v}=v\hat{\bf y}$ and $\delta =0$); and magnetosonic waves, for
which the velocity of the fluid makes an arbitrary angle with the
background field (${\bf k}=k\hat{\bf x}, {\bf B}_0= B_x\hat{\bf
x}+B_y\hat{\bf y}, {\bf b}=b\hat{\bf y}, {\bf v}=v_x\hat{\bf x} +
v_y\hat{\bf y}$). Note that sound waves, which propagate along the
background field without affecting it (${\bf B}_0\,\|\, {\bf k}\,\|\,
{\bf v}$ and ${\bf b} =0$), are a special case of magnetosonic
waves. In all dispersion relations and their solutions, $\theta$
denotes the angle between the background magnetic
field and the wave vector.

The dispersion relation for magnetosonic modes allows two solutions:
slow magnetosonic modes and fast magnetosonic modes. Fast magnetosonic
modes are similar in nature to sound waves, while slow magnetosonic
modes are closer in character to Alfv\'en waves. This fact plays an
important role in the damping of magnetosonic waves, and is apparent
for weak magnetic fields ($\rho_B \ll \rho_{\rm fluid}$) where fast
magnetosonic modes oscillate with ${\bf v}$ almost along the direction
of ${\bf k}$ and involve oscillating density perturbations, while slow
magnetosonic modes oscillate almost perpendicularly to ${\bf k}$ and
have close to vanishing density perturbations. In the special case
${\bf k}\,\|\,{\bf B}_0$ fast magnetosonic waves become sound waves
and there are no slow magnetosonic solutions. (For a discussion of MHD
modes see, for example, Ref.~\cite{J62}.)

We obtain the dispersion relations by substituting
Eq.~\ref{eq:time_depend} into
Eqs.~\ref{eq:finalfluid0}--\ref{eq:finalfluid3}. The dispersion
relations are derived to first order in $\kappa'$, $\eta'$, $\xi'$,
which corresponds to the lowest non-trivial expansion in powers of
$l_{\rm mfp}/\lambda$ (where $\lambda$ is the wavelength of a
mode). We use the WKB approximation neglecting the time derivatives of
the Fourier amplitudes, $\partial \Phi_{\bf k}(t)/\partial t \ll
\omega$, and the time derivative of the frequency, $\p \omega/\p t \ll
-\omega^2$, which arises from the $\p^2 {\bf v}/\p t^2$ term in
Eq.~\ref{eq:finalfluid1}. This approximation is valid for modes with
oscillation frequencies  much higher than the expansion rate, $
\omega \gg H $.

This procedure for Alfv\'en waves yields:
\begin{equation}
3\omega^3 \kappa'
+ i\omega^2\biggl(1+R + \va^2\biggr)
+ \omega\biggl(\aoa R +3\eta'\biggl(\frac{k}{a}\biggr)^2\biggr)
- i\va^2cos^2\theta\biggl(\frac{k}{a}\biggr)^2=0\ . \label{eq:Alfven}
\end{equation}
Complete dispersion relations for all MHD modes are, due to their
length, placed in Appendix~\ref{sec:diff_app}. 

In the following two sections, we present the solutions for
oscillation frequencies and damping rates derived from the dispersion
relations in the following two limits: the oscillatory limit, when the
solution is oscillatory with an exponentially decaying amplitude; and
the overdamped limit, when the amplitude of modes exponentially
decrease without completing an oscillation.

\subsubsection{Oscillatory Limit}

The solutions to the dispersion relations, $\omega$, generally consist
of a real and an imaginary part, which represent the oscillation
frequency and the damping rate, respectively. In the oscillatory
limit, the dissipative effects are such that the fluid oscillates
many times as it damps, $ {\rm Re}\, \omega\ll {\rm Im}\, \omega$. 
In this case, the dispersion relations can be solved by considering
all the viscosity and heat conductivity terms as perturbations on the
ideal fluid dispersion relation.

The solutions to the dispersion relations will be
conveniently expressed in terms of the speed of sound:

\begin{equation}
v_s = \sqrt{\biggl({\p p\o\p\rho}\biggr)_S}=
\frac{1}{\sqrt{3(1+R)}} \ ,   
\end{equation}
and the relativistic Alfv\'en speed:
\begin{equation}
v_A = \frac{\tilde{B}_0} {\sqrt{1+R+\tilde{B}_0^2}}\ .
\end{equation} 
The relativistic Alfv\'en speed includes the magnetic field energy
density in the denominator, which ensures that for strong magnetic
fields the Alfv\'en speed does not exceed the speed of light.

For all modes in the oscillatory limit we first solve the dispersion
relations for an ideal fluid and then compute the first order
contributions from the dissipative terms. For clarity, all the
solutions in their general form have been placed in
Appendix~\ref{sec:diff_app}. In this and the following section we give
the solutions for each MHD mode in the cosmologically relevant limit
of weak magnetic fields ($\tilde{B}_0 \ll 1$) and negligible redshift
terms ($\dot{a}/a$).

For weak magnetic fields, the leading terms in the frequencies for
fast magnetosonic waves do not depend either on the magnetic field
strength or on the direction of propagation, and are therefore the
same as the frequencies for sound waves:
\begin{equation}
\omega^{\rm FM}_{\rm osc}=\pm v_s\biggl(\frac{k}{a}\biggr)
		+i\biggl(\frac{R^2}{2(1+R)^2}\kappa' +
 		\frac{2}{1+R}\eta' + \frac{3}{2(1+R)}\xi' \biggr)
		\biggl(\frac{k}{a}\biggr)^2 + O(\tilde{B}_0^2).
			\label{eq:fast_diff_damped}
\end{equation}
This reproduces the solution for propagation and damping of sound
waves given in Weinberg (1971). Similarly, the frequencies of slow
magnetosonic waves and Alfv\'en waves are identical to leading order
in $B_0$ and have the following form:

\begin{equation}
\omega^{\rm SM,A}_{\rm osc}=\pm v_A \cos\theta \biggl(\frac{k}{a}\biggr) + \frac{3}
{2}i{{\eta^{\prime}}\over {(1+R)}} \biggl(\frac{k}{a}\biggr)^2 + O(\tilde{B}_0^2)\
.\label{eq:freq_Alfven}
\end{equation}

The solutions show that, while for small magnetic fields the damping
of slow magnetosonic and Alfv\'en waves proceeds through shear
viscosity, fast magnetosonic waves are damped by shear and bulk
viscosity, as well as heat conductivity. Furthermore, fast
magnetosonic waves damp differently in different regimes: they damp
predominantly by heat conductivity when the matter density is larger
than the radiation density, and by shear viscosity when the radiation
density dominates. Note that after taking the time dependence of all
variables into account, the expansion of the fluid affects the
frequencies directly through $\dot{a}/a$ terms (see
Appendix~\ref{sec:diff_app}), and indirectly through the integral in
Eq.~\ref{eq:time_depend}. For instance, as in the case of sound waves
when $R \ll 1$, the oscillation frequency of a fast magnetosonic mode
with a given wavelength in a radiation dominated expansion is twice
the frequency of the same wavelength mode in a static background
metric.

\subsubsection{Overdamped Limit}

When dissipative effects become very strong, oscillations of MHD modes
are inhibited and the evolution of a given MHD mode is dominated by
the exponential decay of its amplitude with time. We seek solutions in
the extremely overdamped regime by expanding the equations in powers
of ${\rm Re}\,\omega_{\rm osc} / {\rm Im}\,\omega_{\rm osc}$, where
$\omega_{\rm osc}$ is the frequency of a wave derived in the
oscillatory limit.

In general, a dispersion relation expanded in powers of ${\rm
Re}\,\omega_{\rm osc} / {\rm Im}\,\omega_{\rm osc}$ has several
solutions distinguished in nature by their initial conditions.  For
example, in the case of Alfv\'en waves, fast decaying solutions arise
from initial conditions such that when the velocities of the fluid are
damped away by shear viscosities the amplitude of the magnetic
perturbations vanish as well. In contrast, when initial conditions
generate slowly decaying Alfv\'en modes, the fluctuations are not
erased as the velocities damp to zero; after the damping of fluid
motions, the remaining magnetic forces tend to accelerate the fluid,
although inefficiently because of the strong viscous damping. Since
energy dissipation rates are proportional to the peculiar fluid
velocity, the timescale for dissipation of the magnetic field
perturbation of slow decaying modes may be extremely large.
 
While the amplitudes of the fast decaying modes damp at rates similar
to the ones calculated in the oscillatory regime (\S 2.3.1), the
amplitudes of slowly decaying modes decay at significantly different
rates.  For weak magnetic fields ($\tilde{B}_0 \ll 1$), the decay rate
for the amplitude of overdamped slow magnetosonic modes is
\begin{equation}
\omega^{\rm SM}_{\rm od} = i \kappa^{\prime} v_A^2 \biggl(\frac{k}{a}\biggr)^2 
			   + i \frac{v_A^2 \cos^2\theta}
				{3 \eta^{\prime}}\, ,  
				\label{eq:slow_diff_overd}
\end{equation}
and the decay rate for Alfv\'en modes is
\begin{equation}
\omega^{\rm A}_{\rm od} = i \frac{v_A^2 \cos^2\theta}{3 \eta^{\prime}}\, .  
\label{eq:alf_diff_overd}
\end{equation}                           

Note that all modes with relativistic propagation velocities (${\rm
Re}\,\omega_{\rm osc} \sim k/a$) never enter the overdamped regime in
the diffusion limit. For this reason a discussion of overdamped
relativistic sound and fast magnetosonic waves is not necessary.

\section{Damping of Magnetohydrodynamic Modes in the Radiation 
Free-Streaming Limit}

Slow magnetosonic and Alfv\'en modes which become overdamped during
the diffusion regime survive the damping and with the expansion of the
universe enter the free-streaming regime when the mean free path of
the decoupling particles grows to be larger than the wavelength of a
mode. In order to investigate the additional damping that these modes
undergo in the free-streaming regime, we study the general case of MHD
in an expanding fluid in the presence of a uniform background. Similar
to our analysis of the diffusion regime, we study the evolution of a
single dissipative fluid. However, in this case the fluid is comprised
of all the particles with mean free paths much shorter than the
wavelength of the MHD mode, while the decoupling particle species,
whose mean free path is now large as it decouples from the rest of the
fluid, represents a uniform background on the scales of interest. The
dissipation arises from occasional collisions of the fluid particles
with the relativistic background.

We generically define a drag coefficient $\alpha$
and a heat exchange coefficient $\gamma$ in the following way: the drag
force per unit volume on the fluid element from scattering with the
background particles is given by
\begin{equation}
{\bf f}\equiv -\alpha{\bf v}\rho_{\rm fluid}\ ,\label{eq:101}
\end{equation}
and the heat exchanged between the fluid element and the background is 
\begin{equation}
\frac{\partial \rho_{\rm thermal}}{\partial t}\equiv -\gamma{T_1\o T_0}
\rho_{\rm thermal}  \ .\label{eq:102}
\end{equation}
The exact form of these coefficients is obtained by calculating the
transfer of momentum and heat per scattering and averaging it over the
distribution of background and fluid particles. We presented the
coefficients later separately for neutrino decoupling and photon
decoupling.

In order to derive the free-streaming fluid equations, we use the
fluid equations developed in \S 2, as well as techniques for finding
solutions described therein. The heating rate from Eq.~\ref{eq:102} is
incorporated into Eq.~\ref{eq:general_energy}, while the drag force
from Eq.~\ref{eq:101} is added to Eq.~\ref{eq:general_momentum}
together with the magnetic field contribution from
Eq.~\ref{eq:mag_momentum}. Although the local thermodynamic
equilibrium between the fluid and the free-streaming component does
not hold in general, we consider the case in which the mean scattering
time between particles of the fluid component and the free-streaming
component is shorter than the characteristic expansion time scale of
the fluid. In this case the temperature and velocity of the background
is the same as the average temperature and velocity of the fluid. All
other assumptions, including non-relativistic fluid velocities, are
carried over from \S 2.

The resulting equations are:
\begin{equation}
{\p\rho_1\o\p t}+ {1\o a}(\rho_0+p_0){\bf\nabla\cdot v}+3\aoa(\rho_1
+p_1)=-\gamma\frac{T_1}{T_0}\rho_{\rm thermal}\ ,\label{eq:105}
\end{equation}

\begin{equation}
\frac{1}{a^4} {\p\o\p t}(a^4(\rho_0+p_0){\bf v})
+\frac{1}{a}{\bf\nabla}p_1 + {1\o 4\pi}\tilde{\bf
B}_0{\bf\times}\bigl(\frac{\p {\bf v}}{\p t} \times\tilde{B}_0\bigr)+{1\o
4\pi}\tilde{\bf
B}_0{\bf\times}\bigl({\bf\nabla\times\tilde{b}}\bigr)=-\alpha {\bf v}
\rho_0\ ,\label{eq:106}
\end{equation}

\begin{equation}
{\p n^b_1\o\p t}+3\aoa n^b_1+{n^b_0\o a}{\bf\nabla\cdot v}=0\
,\label{eq:107}
\end{equation}
and together with Maxwell's equations Eq.~\ref{eq:finalfluid2} -
Eq.~\ref{eq:finalfluid4} they form a complete set.

To derive dispersion relations for a given fluid, we have to specify
the energy density, matter density, and pressure, and substitute these
into the above set of equations. This is done in the rest of \S 3 for
two fluid combinations: a baryonic fluid with free-streaming photons;
and a fluid which consists of baryons and relativistic particles like
photons and $e^+ e^-$ pairs, in a background of free-streaming
neutrinos. All the dispersion relations as well as the solutions for
fast magnetosonic modes are given in the appendices. Here we present
the solutions to dispersion relations for slow magnetosonic modes and
Alfv\'en modes, the two modes that in the presence of weak magnetic
fields survive into the free-streaming regime before recombination.

\subsection{Neutrino Free-Streaming Limit}

Around neutrino decoupling the fluid consists of tightly coupled
baryons, photons, and $e^+ e^-$ pairs. The dominant component of the
pressure is radiation pressure
\begin{equation}
p_r=\frac{1}{3} \rho_r , 
\end{equation}
and, since $R\simeq 0$, the speed of sound is
\begin{equation}
v_s = \frac{1}{\sqrt{3}}\ .
\end{equation}
The energy density (equal to $\rho_{\rm thermal}$) has
contributions from all relativistic particles, counted in the number
of degrees of freedom $g_r$,
\begin{equation}
 \rho_r = g_r\frac{\pi^2}{30}T^4. 
\end{equation}
The heat exchange coefficient and drag coefficient defined by
Eq.~\ref{eq:102} and Eq.~\ref{eq:101} may be computed
by averaging the transfer of energy in each scattering between a
distribution of background particles and a distribution of fluid particles
\cite{JF}. For neutrino free-streaming
they have the form:
\begin{equation}
\gamma \simeq \sigma_{w} n_{w} \frac{\rho_\nu}{\rho_\gamma} = 
		\frac{g_{\nu}}{g_r l_{\nu}} \, ,
\label{eq:nu_alpha}
\end{equation}
and
\begin{equation}
\alpha \simeq \gamma\, ,
\label{eq:nu_gamma}
\end{equation}
where the second equation follows on dinensional grounds.
Here $\sigma_{w}$ is the cross section for scattering of neutrinos
with other weakly interacting particles, $n_{w}$ is the number density
of weakly interacting particles (scatterers), $g_{\nu}$ is the neutrino statistical
weigth, and $l_{\nu}$ is the neutrino mean free path.

Following the steps used in the diffusion regime (\S 2.3), we obtain
dispersion relations for the different MHD modes and present them in
Appendix~\ref{sec:nu_fs_app}.

\subsubsection{Oscillatory Limit}

The oscillation frequencies and damping rates for slow magnetosonic
and Alfv\'en modes are again obtained by first solving the dispersion
relations for an ideal fluid and then solving for the first order
dissipative terms. In terms of the previously defined Alfv\'en speed,
the solutions for small magnetic fields are the same for slow
magnetosonic and Alfv\'en modes and have the form:
\begin{equation}
\omega^{\rm SM,A}_{\rm osc} = \pm v_A \cos\theta \biggl(\frac{k}{a}\biggr) +
\frac{3i}{8}\alpha \, . 
\end{equation}
The frequencies for oscillatory fast magnetosonic waves (including
sound waves) are presented in Appendix~\ref{sec:nu_fs_app}.

\subsubsection{Overdamped Limit}

As in the diffusion regime, the solutions in the extremely overdamped
regime (${\rm Re}\, \omega_{\rm osc} \ll {\rm Im}\, \omega_{\rm osc})$
are derived by expanding the equations in powers of ${\rm Re}\,
\omega_{\rm osc} / {\rm Im}\, \omega_{\rm osc}$. The overdamped
solution in the case of weak magnetic fields is, for slow
magnetosonic modes
\begin{equation}
\omega^{\rm SM}_{\rm od} = \frac{i}{4}\gamma v_A^2 + 
\frac{4i v_A^2 \cos^2\theta}{3\alpha} \biggl(\frac{k}{a}\biggr)^2,
\label{eq:1400}
\end{equation}
and for Alfv\'en waves:
\begin{equation}
\omega^{\rm A}_{\rm od} = \frac{4i v_A^2 \cos^2\theta}{3\alpha} 
\biggl(\frac{k}{a}\biggr)^2
\,. \label{eq:1401}
\end{equation}

As in the case of radiation diffusion (\S 2.3), relativistic fast
magnetosonic modes do not become overdamped in the free-streaming
regime.

\subsection{Photon Free-Streaming Limit}

Fluid equations for the modes in the free-streaming limit around
photon decoupling are somewhat different from the cases analyzed
so far. Namely, the only contribution to the energy density of the
perturbations is the thermal energy density of baryons,
\begin{equation}
\rho_{\rm thermal} = \frac{3}{2}(n_e+n_p)T = 3 n_b T,
\end{equation} 
which enters Eq.~\ref{eq:105} but can be neglected in Eq.~\ref{eq:106}
because it is much smaller than the matter density. Here $n_e$ and
$n_p$ are electron and proton number densities respectively.
Furthermore, since the photons can be considered decoupled on the
free-streaming scales, the only pressure left to support the
oscillations is the pressure of the baryonic fluid itself:
\begin{equation}
p_b = (n_e + n_p) T = 2 n_b T.
\end{equation}
This yields the gradient of pressure in Eq.~\ref{eq:106} which depends
both on density and temperature fluctuations and is best expressed
through the sound speed:
\begin{equation}
\frac{1}{\rho} {\bf \nabla}p = \frac{3}{5} c_s^2 ({\bf \nabla}\Delta +
{\bf \nabla}\delta),
\end{equation}
where $c_s$ is the adiabatic baryonic speed of sound for a fully
ionized proton-electron fluid
\begin{equation}
c_s=\sqrt{\biggl({\p p_b\o\p\rho_b}\biggr)_S}= \sqrt{{10\o 3}{T\o
m_p}}\ . \label{eq:108}
\end{equation}

With these substitutions for the densities and the pressure, we obtain
dispersion relations for different MHD modes and present them in
Appendix~\ref{sec:ph_fs_app}.

The drag and heat exchange coefficients, which appear in the
dispersion relations and their solutions, are similarly obtained as in
the neutrino free-streaming case, and have the following form \cite{P65}:
\begin{equation}
\alpha \simeq \sigma_{\rm T} n_e \frac{\rho_\gamma}{\rho_b} = 
			\frac{1}{l_\gamma R},
\label{eq:ph_alpha}
\end{equation}
and
\begin{equation}
\gamma \simeq \frac{m_p}{m_e} \alpha.
\label{eq:ph_gamma}
\end{equation}

\subsubsection{Oscillatory Limit}

Unlike in the relativistic cases when the photon pressure dominates,
the structure of the non-relativistic equations with free-streaming
photons allows for oscillating magnetosonic modes with two different
propagation velocities and damping rates. These modes are commonly
referred to as adiabatic and isothermal, depending if heat transport is
slow or rapid compared to the oscillation time: a mode is adiabatic
when $\omega \gg \gamma$ and $\omega \gg \alpha$, and it is isothermal
when $\gamma \gg \omega \gg \alpha$. Alfv\'en modes have only one
solution since they do not include density or temperature fluctuations
and therefore are not affected by heat transport.

Again we derive the dispersion relations from
Eq.~\ref{eq:finalfluid2}--\ref{eq:finalfluid4} and
Eq.~\ref{eq:105}--\ref{eq:107} and place it in Appendix~\ref{sec:ph_fs_app}. Before
presenting the solutions for slow magnetosonic and Alfv\'en modes, it
is useful to introduce the non-relativistic Alfv\'en speed,
\begin{equation}
c_{A} = \frac{\tilde{B}_0} {\sqrt{R}}\ .
\end{equation} 

The oscillation frequency and the damping rate for slow magnetosonic
waves in the adiabatic regime ${\rm Re}\,\omega\gg\gamma,\alpha$ are
given by:
\begin{equation}
	\omega^{\rm SM}_{\rm osc}=\pm c_{A} \cos\theta \biggl(\frac{k}{a}\biggr) 
		+i\left({\alpha\o 2}+ {\gamma \o 5}{{\tilde
		B}_0 \sin^2 \theta \o {c^2_s R}}\right) \, ,
			\label{eq:116}
\end{equation}
and
\begin{equation}
	\omega^{\rm SM}_{\rm osc}=\pm c_s \cos\theta \biggl(\frac{k}{a}\biggr)
		+i\left({\alpha\o 2} + {\gamma \o 5}\right)\, ,
			\label{eq:117}
\end{equation}
where the upper solution corresponds to weak magnetic fields such that
$c_s\gg c_A$ and the lower solution to strong magnetic fields with
$c_s\ll c_A$. It is important to remember that both of these solutions
are derived for a background magnetic field whose energy density is
much smaller than the energy density in photons. The condition for
adiabaticity is dependent on the strength of the magnetic field since
oscillation frequencies of magnetosonic waves are different for strong
and weak magnetic fields. Slow magnetosonic modes in the isothermal
regime, $\gamma\gg {\rm Re}\,\omega\gg\alpha$, have the solution

\begin{equation}
	\omega^{\rm SM}_{\rm osc}=\pm c_{A} \cos\theta \biggl(\frac{k}{a}\biggr) 
				+i{\alpha\o 2},\label{eq:118}
\end{equation}
and
\begin{equation}
	\omega^{\rm SM}_{\rm osc}=\pm \sqrt{3\o 5} c_s \cos\theta \biggl(\frac{k}{a}\biggr)
				+i{\alpha\o 2},\label{eq:119}
\end{equation}
where again the upper solution is for $c_s\gg c_A$ and the
lower solution is for $c_s\ll c_A$. 
  
The result for Alfv\'en waves is
\begin{equation}
	\omega^{\rm A}_{\rm osc} =\pm c_{A} \cos\theta \biggl(\frac{k}{a}\biggr)
			+i \frac{\alpha}{2}.\label{eq:120}
\end{equation}

The frequency of non-relativistic fast magnetosonic waves in the
adiabatic and isothermal limits are placed in
Appendix~\ref{sec:ph_fs_app}.

\subsubsection{Overdamped Limit}

The frequencies for slow magnetosonic waves corresponding to slow
exponential decay are:
\begin{equation}
\omega^{\rm SM}_{\rm od}=i{c_A^2 \cos^2\theta \o \alpha} 
\biggl(\frac{k}{a}\biggr)^2\ , \label{eq:1300}  
\end{equation}
and
\begin{equation}
\omega^{\rm SM}_{\rm od}=i{3c_s^2 \cos^2\theta \o 5\alpha} 
\biggl(\frac{k}{a}\biggr)^2\ , \label{eq:1301}
\end{equation}
with the upper solution for $c_A\ll c_s$, and the lower solution for $c_A
\gg c_s$. The slowly decaying mode of overdamped Alfv\'en waves has
frequency
\begin{equation}
	\omega^{\rm A}_{\rm od}=i{c_A^2 \cos^2\theta \o \alpha} \biggl(\frac{k}{a}\biggr)^2\ .
		\label{eq:130}
\end{equation}

In contrast to relativistic MHD, some non-relativistic fast
magnetosonic modes enter the overdamped regime. Their damping rates
can also be found in Appendix~\ref{sec:ph_fs_app}.

\section{Damping of Magnetic Fields in the Early Universe}

In this section we discuss the implications of the damping of
magneto-hydrodynamic modes for the evolution of cosmological magnetic
fields. Magnetic fields generated in the early universe are likely to
be randomly oriented, spatially varying fields with small coherence
lengths, usually of the order of the horizon at the epoch when the
fields were created. We assume that the magnetic fields are created
with magnetic field energy below equipartition with the radiation
energy density, i.e. $\tilde{B}_0\ll 1$. 
For an arbitrary magnetic field configuration, we
choose a separation of scales such that in a given volume the field
can be described as an approximately force-free background magnetic
field ${\bf B}_0$, and a spectrum of propagating modes ${\bf b}({\bf
k})$, where $|{\bf b}| \ll |{\bf B}_0|$. In this case we can decompose
the propagating modes into slow and fast magnetosonic, and Alfv\'en
modes with different wave vectors $\bf k$ and different phases.
Although the condition $|{\bf b}| \ll |{\bf
B}_0|$ may not be easily achieved for every field configuration, the
predicted evolution of propagating modes is indicative of the general
field evolution. In particular, the efficient viscous damping
discussed in this paper should cause the dissipation of magnetic energy in
generic field configurations.

We are interested in the evolution of individual MHD modes from before
the epoch of neutrino decoupling to recombination. For each epoch we
wish to determine the characteristic scales over which pre-existing
cosmic magnetic fields are damped. As previously described, the
evolution of fast magnetosonic waves is distinctively different from
the evolution of slow magnetosonic modes and Alfv\'en waves; therefore
they are discussed separately: fast magnetosonic waves in \S 4.1, and
slow magnetosonic and Alfv\'en waves in \S 4.2.

For the calculation of the damping scales we need the expressions for
the mean free path of the decoupling particles as well as the ratio of
the baryon density to the photon density. While the universe cools
from temperatures below the electroweak breaking scale ($T \sim$ 100
GeV) to neutrino decoupling ($T_{\nu} \sim$ MeV), neutrinos are the
particles with the longest mean free path and therefore the most
efficient momentum and heat transporters. The neutrino mean free path
at temperature $T$ can be written as
\begin{equation} 
	l_{\nu}(T)\simeq {1\o G_F^2T^2(n_l+n_q)}
	\simeq 10^{11} {\rm cm} \,\biggl({T\over{\rm MeV}}\biggr)^{-5}
		\biggl(\frac{g_l+g_q}{8.75}\biggr)^{-1}\,, 
			\label{eq:nu}
\end{equation}  
where $n_l$ and $n_q$ are lepton and quark number
densities, $G_F$ is Fermi's constant, and $G_F^2T^2$ is a typical
weak interaction cross section.  The quantities $g_l$ and
$g_q$ are the statistical weights of relativistic weakly interacting
leptons and quarks present at the epoch of interest. 

At temperatures below the completion of the $e^+e^-$-annihilation
($T\sim 20$ keV), heat and momentum are most efficiently transported
by photons. The dominant process limiting the photon mean free path
during this period is Thomson scattering of photons off electrons
which gives the photon mean free path the following form:
\begin{equation}
	l_{\gamma} (T) \simeq {1\o \sigma_{T}n_e}
	\simeq 10^{22} {\rm cm} \, 
		\biggl({T\over 0.25{\rm eV}}\biggr)^{-3}
		\biggl({\Omega_b h^2\over 0.0125}\biggr)^{-1} 
		X_e(T)^{-1}\,, \label{eq:gamma}
\end{equation}
where $X_e$ is the number of free electrons per baryon, $\sigma_{T}$
is the Thomson cross section, and $n_e$ is the electron density.

The damping of MHD waves is particularly efficient during the epochs
of neutrino decoupling and recombination, when efficient momentum
transfer and heat transport arise from the quickly growing mean free
path of the decoupling particles. Therefore, all the variables in the
above equations have been scaled to their values at those
epochs. Also, since at neutrino decoupling the dominant scattering
process is scattering of neutrinos off leptons, the appropriate values
for the parameters in Eq.~\ref{eq:nu} are $g_l=8.75$ ($e^{\pm}$ and
six neutrino species) and $g_q=0$. The remaining unspecified
parameter, the ionization fraction $X_e$, drops 
within a short time from 1 to $\sim
10^{-5}$ during recombination which occurs approximately at
$T_{\gamma}^d \simeq 0.25$ eV.

The baryon mass density is negligible when compared to the radiation
energy density during neutrino decoupling ($R\approx 0$), while around
recombination it is approximately given by
\begin{equation}
	R={3\rho_b\o 4\rho_r}
	\approx 0.4 \biggl({T\o 0.25 {\rm eV}}\biggr)^{-1}
		\biggl({\Omega_b h^2\o 0.0125}\biggr), 
			\label{eq:22}
\end{equation} 
where  $\Omega_b$ is the fractional contribution of
baryons to the closure density and $h$ is the present Hubble
constant in units of 100 km s$^{-1}$ Mpc$^{-1}$. 
In writing Eq.~\ref{eq:22} we implicitly assume that neutrinos have
decoupled from the remaining particle species.

\subsection{Damping of Fast Magnetosonic Waves in the Early Universe}

The damping of all fast magnetosonic modes is to leading order
equivalent to the damping of sound waves if the energy density in the
large-scale magnetic field is much smaller than the energy density in
radiation. The damping occurs due to the diffusion of either neutrinos
prior to neutrino decoupling, or photons before recombination. From
the expressions in Eqs.~\ref{eq:nu} and \ref{eq:gamma}, which
represent mean free paths in proper units, it can be seen that in both
cases the comoving mean free path grows with the expansion of the
universe. As a consequence, MHD modes with wavelengths $\lambda_p = 2
\pi a(T)/k$ smaller than the mean free path $l_{\rm mfp} (T)$ at time
$t(T)$ were in the diffusion regime at some prior time in the early
universe, i.e. $\lambda_p(T') \gg l_{\rm mfp} (T')$ at $t(T')$ .

The amount of damping that fast magnetosonic modes undergo in the
diffusion regime can be calculated using the damping rates in
Eq.~\ref{eq:fast_diff_damped}. From the leading damping term, $2i
\eta^{\prime} (k/a)^2/(1+R)$, and the definitions in Eqs.~\ref{eq:9}
and \ref{eq:shear}, we see that the amplitude of the perturbation is
damped between time $t=0$ and $t$ by
\begin{equation}
\exp\biggl(-\int_0^t\, \frac{l_{\rm mfp}}{\lambda_p^2} 
		{\rm d}t^{\prime}\biggr).
\end{equation}
If we define a characteristic damping scale as the largest comoving
wavelength of an MHD mode whose initial amplitude has been damped by
at least one e-folding by time $t$, this damping scale approximately
corresponds to the comoving photon/neutrino diffusion length,
\begin{equation}
	d^2 = \int_0^t {l_{\rm mfp}(t^\prime)\over
		a^2(t^\prime)} {\rm d}t^\prime,
\end{equation}
which is the distance a photon/neutrino has random walked between time
$t=0$ and $t$. 

In this section we present all the damping scales in comoving units,
in particular, comoving to the present epoch,
unless stated otherwise. The diffusion damping scale calculated for
fast magnetosonic waves prior to neutrino decoupling ($T \geq 1$ MeV)
is:
\begin{equation}
\lambda^{\rm FM}_{\nu}\simeq 2\times 10^{20} {\rm cm} \,
              \biggl({T\o {\rm MeV}}\biggr)^{-{5\o 2}} 
              \biggl({g_{r}\o 10.75}\biggr)^{-{3\o 4}}
              \biggl({g_{\nu}\o 5.25}\biggr)^{1\o 2}
              \biggl({g_l+g_q\o 8.75}\biggr)^{-{1\o 2}}
              \, .
              \label{eq:30}
\end{equation}
Note that the damping scale at neutrino decoupling converted to
proper units approximately corresponds to the causal horizon at that
time, $\lambda_\nu ({\rm MeV}) \approx 5\times 10^{10} {\rm cm}\,$.

In a similar fashion we can compute the comoving damping scale of fast
magnetosonic waves due to the effects of the finite photon mean free
path at lower temperatures. Around recombination, where we assume
matter domination, this damping scale is
\begin{equation} 
\lambda^{\rm FM}_{\gamma}\simeq 7\times 10^{25} \, {\rm cm} 
                  \biggl({T\o {\rm 0.25 eV}}\biggr)^{-{5\o 4}} 
                  \biggl({\Omega_b h^2\o 0.0125}\biggr)^{-{1\o 2}}
                  \bigl(\Omega_0h^2\bigr)^{-{1\o 4}}\,.
              \label{eq:31} 
\end{equation} 
In this expression $\Omega_0$ is the total density in units of the critical
density at the present epoch.

Baryonic mass scales can be associated with the damping length scales
by defining
\begin{equation} 
M \equiv {4\pi\o 3}\rho_b(T)\biggl({\lambda(T) a(T)\o 2}\biggr)^3\ ,
\end{equation} 
where $\rho_b$ is the average baryon mass density at temperature
$T$. The baryonic mass scale associated with the damping
scale of fast magnetosonic waves by neutrino diffusion
around neutrino decoupling is
\begin{equation}
M^{\rm FM}_{\nu}\simeq 10^{-4} {\rm M}_{\odot} \,
              \biggl({T_{\nu} \o {\rm MeV}}\biggr)^{-{15\o 2}} 
              \biggl({g_{r}\o 10.75}\biggr)^{-{9\o 4}}
              \biggl({g_{\nu}\o 5.25}\biggr)^{3\o 2}
              \biggl({g_l+g_q\o 8.75}\biggr)^{-{3\o 2}}
              \biggl({\Omega_bh^2\o 0.0125}\biggr) \, ,
\label{eq:32}
\end{equation}
and by photon diffusion around recombination
\begin{equation}
M^{\rm FM}_{\gamma}\simeq  10^{13} {\rm M}_{\odot} \, 
                  \biggl({T_{\gamma} \o {\rm 0.25 eV}}\biggr)^{-{15\o 4}} 
                  \biggl({\Omega_b h^2\o 0.0125}\biggr)^{-{1\o 2}}
                  \bigl(\Omega_0h^2\bigr)^{-{3\o 4}} \, .
\label{eq:33} 
\end{equation} 

The above calculations for the damping scales are approximate in that
we assume that the damped modes are in the diffusion regime,
$l_{\nu,\gamma}\ll \lambda_p$. The diffusion approximation is not a
valid approximation late within the decoupling epochs. Around neutrino
decoupling, while our analysis uncovers the order of magnitude of the
damping mass scale, an improvement on the diffusion approximation
only, would not yield more accurate results since the calculated
damping scale approximately corresponds to the causal horizon at that
time. Around recombination, our treatment is analogous to the
calculations of the damping of sound waves in Ref.~\cite{W71}; in
writing Eq.~\ref{eq:31} we assume an instantaneous recombination while
the ionization fraction $X_e$ decreases gradually to zero during
recombination. More detailed treatments for the damping of sound waves
during recombination have been performed using the Boltzman equation
\cite{PY,P81,JW} or using a two-fluid model
\cite{BL,PV}. Since the dominant damping terms of fast
magnetosonic waves are the same as those of sound waves, values for
the sound wave damping scale calculated in more accurate models may be
used as better approximations to the fast magnetosonic damping scale.
A review of previous Silk scale calculations is given in Ref.~\cite{JW}.

\subsection{Damping of Slow Magnetosonic and Alfv\'en Waves in the Early
Universe}

Unlike the damping of fast magnetosonic waves, the damping of slow
magnetosonic and Alfv\'en waves in the early universe proceeds through
several different stages. We illustrate these stages by following an
Alfv\'en wave from after neutrino decoupling up to recombination.

Initially, in the diffusion regime where $\lambda \gg l_{\rm mfp}$, a
wave oscillates and damps in the same manner as described for fast
magnetosonic waves. The oscillation frequency and the damping rate are
shown in Eq.~\ref{eq:freq_Alfven}, and after using Eqs.~\ref{eq:9} and
\ref{eq:shear} they approximately become
\begin{equation}
	\omega^{\rm A}_{\rm osc}\simeq\pm v_A\biggl(\frac{k}{a}\biggr) 
\cos\theta+il_{\gamma}\biggl(\frac{k}{a}\biggr)^2 
		\, .\label{eq:301}
\end{equation}
The damping rate in this expression is the same as the damping rate
for fast magnetosonic modes, and is valid as long as $ v_A\cos\theta\gg
l_{\gamma}k/a$. The crucial difference, however, is that for a small
background magnetic field in the early universe, the oscillation
frequency of an Alfv\'en mode ($v_A k/a$) is much smaller than the
oscillation frequency of a fast magnetosonic mode with the same
wavelength ($v_s k/a$). While all fast magnetosonic modes of interest
satisfy the condition for damping in the oscillatory regime ($v_s \gg
l_{\gamma} k/a$), an Alfv\'en mode can become overdamped when, with
the expansion of the universe, the mean-free-path becomes large enough
for dissipative effects to overcome the oscillation ($v_A cos\theta
\simeq l_{\gamma}k/a$). One may
define a temperature dependent wavelength $\lambda_{\rm od}$,
\begin{equation}
	\lambda_{\rm od}(T,B_0)\simeq {2\pi l_{\gamma}(T) \o 
			v_A\cos\theta}\ ,\label{eq:305}
\end{equation} 
such that modes with proper wavelength $\lambda_p \ge \lambda_{\rm
od}$ are oscillatory while modes with $\lambda_p \le \lambda_{\rm od}$
are overdamped.

Overdamped modes are a superposition of fast and slowly decaying
overdamped modes, and the relative amplitudes depend on the phase of
the mode when it becomes overdamped. Fast decaying overdamped modes
damp at rates similar to the oscillatory modes, and therefore their
damping is equivalent to the damping of fast magnetosonic modes
discussed in the previous section. In this section, we follow the
significantly different evolution of the slowly decaying overdamped
modes which experience the least damping. The amplitude of a slowly
decaying overdamped Alfv\'en mode damps with a rate
given by Eq.~\ref{eq:alf_diff_overd}, which in terms of the photon
mean-free-path is
\begin{equation}
	\omega^{\rm A}_{\rm od}\simeq i{v_A^2\cos^2\theta\o l_{\gamma}}\ .
			\label{eq:302}
\end{equation}
Since the damping rate is
inversely proportional to the growing mean free path, the integrated
damping rate is much smaller than the integrated damping rate of fast
magnetosonic modes during the same period. As a result, the damping
in overdamped diffusion is inefficient and the damping scales of 
Alfv\'en modes at the end of the diffusion regime are
smaller than the damping scales of fast magnetosonic modes.

The left-hand-side of Fig.~\ref{fig:scales-fig} (left of the dotted
line) illustrates the evolution of Alfv\'en waves in the diffusion
regime for $\cos\theta =1$ and a background magnetic field of
$\tilde{B}_0\approx 10^{-3}$ (corresponding to $B_0\approx 3\times
10^{-9}$ Gauss at present). In the diffusion regime there are two
important temperature dependent scales: the photon diffusion length
scale (dashed line), which is the scale damped by one e-folding by the
time the universe cools to temperature $T$ provided that modes are
still in the oscillatory regime; and the overdamping length scale
given by Eq.~\ref{eq:305} (solid line), which shows the temperature at
which a mode with comoving wavelength $\lambda_c$ enters the
overdamped regime. The modes which do not damp significantly in the
radiation diffusion regime before they become overdamped, preserve
their amplitude until they reach the free-streaming regime, apart from
small additional damping during the transition itself. On the graph,
these are all the modes with comoving wavelength larger than that
given by the intersection of the solid line with the dashed
line. Therefore, the intersection of these two lines roughly
represents the largest Alfv\'en mode that is damped by one e-folding
in the diffusion regime. Its position depends on the strength of the
background magnetic field and on the angle between the field and the
wave vector.

Some overdamped modes enter the free-streaming regime 
before recombination (to the right of
the dotted line in Fig.~\ref{fig:scales-fig}) as the mean free path of
the decoupling particles grows with the expansion. The dissipation
coefficients in the free-streaming regime are inversely proportional
to the mean free path unlike those in the diffusion regime. This
implies that, when a wave enters the free-streaming regime, it is
initially overdamped and becomes oscillatory when the drag force and
the heat conduction decrease as the mean free path increases. It also
implies that modes in the free-streaming regime undergo most damping
while overdamped. The damping rates during free-streaming, derived
using Eq.~\ref{eq:ph_alpha} in Eqs.~\ref{eq:120} and \ref{eq:130},
are:
\begin{equation}
\omega^{\rm A}_{\rm od}\simeq i{c_A^2} 
l_{\gamma} \biggl({\rho_b\o\rho_{\gamma}}
	\biggr)\biggl(\frac{k}{a}\biggr)^2 \cos^2\theta\ ,\label{eq:303}
\end{equation}
when a wave is overdamped, and
\begin{equation}
	\omega^{\rm A}_{\rm osc}\simeq 
\pm c_A \biggl(\frac{k}{a}\biggr) \cos\theta
			+i{1\o l_{\gamma}}{\rho_{\gamma}\o\rho_b}.
\label{eq:304}
\end{equation}
when it oscillates.

Using Eq.~\ref{eq:303} in Eq.~\ref{eq:time_depend}, we find that for
Alfv\'en waves during overdamped free-streaming, the largest comoving
wavelength whose amplitude is damped by one e-folding at time $t$ is:
\begin{equation} 
(\lambda_{\gamma}^{\rm A}(t))^2 = \int_0^t c_A^2 \cos^2\theta\, R(t')
	{l_{\gamma}(t^{\prime})\over a^2(t^{\prime})} {\rm
	d}t^{\prime} = c_A^2 \cos^2\theta\, d_{\gamma}^2(t) R(t)\ ,\label{eq:904}
\end{equation}
where $d_{\gamma} (t)$ is the comoving photon diffusion length. As we
can see, the damping depends on the strength of the background
magnetic field and the angle between the field and the wave
vector. The above damping length evaluated at recombination is the
characteristic Alfv\'en wave damping length for the free-streaming regime, because
all modes stay overdamped before recombination regardless of the
strength of the magnetic field.

The damping of Alfv\'en waves with free-streaming photons is
illustrated on the right-hand side of Fig.~\ref{fig:scales-fig}. The
dashed line shows the damping scale from Eq.~\ref{eq:904}. The solid
line marks the transition from overdamped to oscillatory behavior,
defined in the same fashion as in the diffusion regime
(Eq.~\ref{eq:305}), with the overdamped region to the left of the
line. Since all the modes that cross the dashed line before
recombination get damped by one e-folding during overdamped
free-streaming, the length scale marked by this line at recombination
represents the free-streaming damping scale for Alfv\'en modes.

In our analysis we have assumed the WKB approximation.
This approximation does not formally hold during all the discussed
epochs for every mode. In particular, the approximation breaks down
for comoving wavelengths between the damping length scale for Alfv\'en
waves in the photon free-streaming regime, $\lambda_{\gamma}^{\rm A}$,
and the photon mean free path (i.e., between the dashed and dotted
lines on the right-hand-side of Fig.~\ref{fig:WKBfigure}). 
Although the damping rates
calculated in the preceding section predict no damping in this region,
some damping is in principle possible since here WKB is not a good
approximation. In Fig.~\ref{fig:WKBfigure} we present the result of a
numerical integration of the magnetic field amplitude of Alfv\'en
waves in different wavelengths through the epoch of interest. The mode
with the largest wavelength which still damps before the epoch of
recombination ($T\simeq 0.25$eV) is consistent with our analytic
estimate. Note that in this calculation we assume that recombination
never occurs (i.e., $X_e \simeq 1$ for $T\le 0.25$eV in
Fig.~\ref{fig:WKBfigure}). We have extended the calculation into this
non-physical regime to learn when the results obtained by the WKB
approximation deviate significantly from the numerical results. From
the plot we see that significant discrepancies happen only with modes
that would damp after recombination if $X_e$ was kept fixed.

The damping of slow magnetosonic waves proceeds similarly to the
illustrated damping of Alfv\'en waves except for two
differences. First, slow magnetosonic waves in overdamped diffusion
damp at a slightly different rate than Alfv\'en waves because of the
additional damping expressed through the extra term in
Eq.~\ref{eq:slow_diff_overd}. This damping is relevant only for the
waves that propagate at very large angles with respect to the
background magnetic field. Second, the damping rates for slow
magnetosonic waves during free-streaming depend on whether the
non-relativistic sound speed is larger or smaller than the
non-relativistic Alfv\'en velocity
(Eq.~\ref{eq:1300}--Eq.~\ref{eq:1301}). The damping scale when $c_s >
c_A$ is the same as the damping scale for Alfv\'en waves, while the
damping scale for $c_s < c_A$ is unique to slow magnetosonic modes.

We can define damping length and mass scales for Alfv\'en and slow
magnetosonic waves analogously to the previous section, and consider
that the waves below these scales would have dissipated by the time of
neutrino/photon decoupling. The general evolution of all the damping
length scales with temperature, followed through the different damping
stages up to decoupling, can be found in
Appendix~\ref{sec:scales_app}. In this section we present the final
damping lengths at neutrino decoupling ($T \simeq 1 {\rm MeV}$) and at
recombination ($T \simeq 0.25 {\rm eV}$), for the background magnetic
field below $3\times 10^{-9}$ Gauss, which is the current estimate of
the upper limit on magnetic fields on Mpc scales (see,
e.g. Ref.~\cite{K94}).

The final damping scale of Alfv\'en waves for this range of the
background magnetic field strengths is determined by the
free-streaming damping length. At neutrino decoupling, most modes in
free-streaming are damped while overdamped, although the largest modes
are damped just as they begin to oscillate again. This determines the
comoving damping scale at neutrino decoupling to be:
\begin{equation}
\lambda_{\nu}^{\rm A} \simeq 10^{17} {\rm cm} \, 
		B_9 \cos\theta 
		\biggl({g_{r}\o 10.75}\biggr)^{-1}
		\biggl({g_{\nu}\o 5.25}\biggr)^{1\o 3}
		\biggl({g_l+g_q\o 8.75}\biggr)^{1\o 3} \, .
		\label{eq:Alfven-length-nu}
\end{equation}
The background magnetic field is expressed through $B_9 = B_0/(3\times
10^{-9} {\rm Gauss})$, where $B_0$ is the background magnetic field
strength scaled to the present epoch. During photon decoupling, the
largest Alfv\'en mode damped by one e-folding during free-streaming is
damped while still in the overdamped regime, and its comoving wavelength is:
\begin{equation}
\lambda_{\gamma}^{\rm A} \simeq 2\times 10^{23} {\rm cm} \, B_9 \cos\theta
		\biggl(\Omega_0 h^2\biggr)^{-{1\o 4}}
              	\biggl(\frac{\Omega_b h^2}{0.0125}\biggr)^{-{1 \o 2}}\, .
			\label{eq:Alfven-length-gamma}
\end{equation}
The baryonic mass scales which correspond to the damping length scales at neutrino and photon decoupling are
\begin{equation}
M_{\nu}^{\rm A} \simeq 10^{-13} {\rm M}_{\odot} \, B_9^3 \cos^3\theta  
              	\biggl(\frac{\Omega_b h^2}{0.0125}\biggr)
              	\biggl({g_{r}\o 10.75}\biggr)^{-3}
              	\biggl({g_{\nu}\o 5.25}\biggr)
		\biggl({g_l+g_q\o 8.75}\biggr)
               	\label{eq:Alfven-mass-nu}
\end{equation}
and
\begin{equation}
M_{\gamma}^{\rm A} \simeq 10^{6} {\rm M}_{\odot}\,B_9^3 \cos^3\theta  
		\biggl(\Omega_0 h^2\biggr)^{-{3\o 4}}
              	\biggl(\frac{\Omega_b h^2}{0.0125}\biggr)^{-{1 \o 2}}\,.
		\label{eq:Alfven-mass-gamma}
\end{equation}

The characteristic damping length scale for slow magnetosonic waves at
neutrino decoupling is the same as the damping length of Alfv\'en
waves (Eq.~\ref{eq:Alfven-length-nu}). At recombination however, the
slow magnetosonic damping scale depends on the value of
the non-relativistic sound speed relative to the non-relativistic Alfv\'en
speed. When $c_s>c_A$ the damping scale for slow magnetosonic waves
in the free-streaming regime is the same as the damping scale for Alfv\'en
waves
(Eq.~\ref{eq:Alfven-length-gamma}). On the other hand, when $c_s<c_A$
the slow magnetosonic damping scale is:
\begin{equation}			
	\lambda_\gamma^{\rm SM} \simeq 3\times 10^{21} {\rm cm} \, \cos\theta
		\biggl(\Omega_0 h^2\biggr)^{-{1\o 4}}\, ,
		\label{eq:slow_length_ph2}
\end{equation}
which is independent of the background magnetic field strength. The
condition $c_s > c_A$ is equivalent to $B_0\simle 5\times 10^{-11}
\,(\Omega_b h^2/0.0125)^{1/2} {\rm Gauss}\,$ for the strength of the
large scale magnetic field at the present epoch, or $\tilde{B}_0
\simle 2\times 10^{-5}\,(\Omega_b h^2/0.0125)^{1/2}$ and $B_9 \simle
1.7\times 10^{-2}\,(\Omega_b h^2/0.0125)^{1/2}$ for the two different
scalings of the magnetic field strength present in our equations.


The baryonic mass scales associated with the damping lengths of slow
magnetosonic modes are:
\begin{equation}
M_{\nu}^{\rm S} \simeq 10^{-13} {\rm M}_{\odot} \, 
		B_9^3 \cos^3\theta  
              	\biggl(\frac{\Omega_b h^2}{0.0125}\biggr)
              	\biggl({g_{r}\o 10.75}\biggr)^{-3}
              	\biggl({g_{\nu}\o 5.25}\biggr)
		\biggl({g_l+g_q\o 8.75}\biggr)\,,
\end{equation}
and
\begin{equation}
M_\gamma^{\rm SM} \simeq \left\{ \begin{array}{ll}
	 \displaystyle 10^{6} {\rm M}_{\odot} \,  
		B_9^3 \cos^3\theta  
		\biggl(\Omega_0 h^2\biggr)^{-{3\o 4}}
              	\biggl(\frac{\Omega_b h^2}{0.0125}\biggr)^{-{1 \o 2}}
		&  \mbox{for $c_s > c_A$} \\ [0.2in]
	\displaystyle 1 {\rm M}_{\odot} \, \cos^3 \theta
		\biggl(\Omega_0 h^2\biggr)^{-{3\o 4}}
              	\biggl(\frac{\Omega_b h^2}{0.0125}\biggr)
		& \mbox{for $c_s < c_A$}
		\label{eq:slow_mass_ph}
\end{array}
\right.
\end{equation}

\section{Conclusions}

In this paper we have studied the effects of dissipation on the
propagation of MHD modes in an expanding fluid composed of matter and
radiation. We have derived the propagation velocities and damping
rates for fast and slow magnetosonic, and Alfv\'en waves in the
radiation diffusion and radiation free-streaming regimes. The derived
damping rates have general applications in magnetized relativistic
and nonrelativistic
astrophysical environments. We have applied the damping rates to the
evolution of MHD modes in the early universe to show that cosmic
magnetic fields suffer significant damping from before neutrino
decoupling to the end of recombination.

Fast magnetosonic waves are damped by radiation diffusion on all
scales smaller than the radiation diffusion length in analogy to the
propagation of sound waves in a demagnetized plasma. The
characteristic damping scales are: the horizon scale at neutrino
decoupling, $M_{\nu}^{\rm FM} \approx 10^{-4} M_{\odot}$ in baryons,
and the Silk mass at recombination, $M_{\gamma}^{\rm FM} \approx
10^{13} M_{\odot}$ in baryons. In contrast to fast magnetosonic waves,
slow magnetosonic and Alfv\'en waves reach an overdamped regime during
which the damping is not very efficient; further significant damping
occurs once the radiation is free-streaming on the scale of the
perturbation. The maximum damping scales for slow magnetosonic and
Alfv\'en modes in general depend on the magnetic field strength and
the direction of propagation with respect to the background magnetic
field. At neutrino decoupling the damping scale is $M_\nu^{\rm A,SM}
\approx 10^{-13} M_{\odot} B_9^3 \cos^3\theta$, the same for both
types of modes. At recombination, if $B_9 < 1.7\times 10^{-2}$, this
scale is $M_\gamma^{\rm A,SM} \approx 10^{6} M_{\odot} B_9^3
\cos^3\theta$, and if $B_9 > 1.7\times 10^{-2}$, the damping scale for
slow magnetosonic modes is different: $M_\gamma^{\rm SM} \approx
1M_{\odot} \cos^3\theta$. The background magnetic field strength
scaled to the present epoch is expressed in terms of $B_9 =
(B_0/3\times 10^{-9} {\rm Gauss})$, the current observational limit on
the large scale magnetic field.

Our findings have various implications for cosmological magnetic
fields. The dissipation of magnetic field energy into heat during the
epoch of neutrino decoupling ensures that most magnetic field
configurations generated prior to neutrino decoupling satisfy big bang
nucleosynthesis constraints. The observed element abundances require
that the energy density in magnetic fields be less than one-third of
the photon energy density during nucleosynthesis \cite{KSV,COST}. Even
if processes prior to neutrino decoupling generate magnetic fields
with initial energy density comparable to the photon energy density,
neutrino damping causes the magnetic energy density to decrease
substantially relative to that of radiation by the time of
nucleosynthesis.

Further dissipation before recombination lessens considerably the
magnetic field energy in primordial magnetic fields available for
generating galactic magnetic fields or density perturbations. The
damping of fast magnetosonic modes implies that a sizable fraction of
the magnetic energy density is erased up to the Silk scale.

Finally, although Alfv\'en and slow magnetosonic modes also undergo
significant damping, their damping scales 
depend on the strength
and the direction of the background magnetic field
and are generally smaller than the
damping scale for fast magnetosonic modes. As long as mode
coupling is not effective, which we expect to be true for
non-oscillating modes before recombination, magnetic energy density
can be stored in Alfv\'en and slow modes on scales well below the Silk
mass. The survival of these modes may be of significance to the
formation of structure on relatively small scales. In particular,
these modes may be responsible for fragmentation of early structures
as well as seeding early star formation. We will further discuss the
consequences of the cosmological evolution of MHD modes in a
subsequent paper.

\section*{Acknowledgments}

We thank B. Berger, J. Geddes, A. Konigl, R. Kulsrud, and  R. Rosner for help
throughout this project. 
We are particularly grateful to B. Chandran for alerting us to the
existence of overdamped magnetohydrodynamic modes. 
This work was performed, in part, under the
auspices of the US Department of Energy by the Lawrence Livermore National
Laboratory under contract number W-7405-ENG-48 and DoE Nuclear Theory
grant SF-ENG-48. It was also supported by DOE at University of
Chicago, DOE and NASA at Fermilab. 
We also acknowledge the hospitality of the Aspen Center 
for Physics where some of this work was performed.

\def\nature{{\rm Nature}} 
\def\nucphys{{\rm Nuc. Phys.}}
\def\nucphysa{{\rm Nuc. Phys. A}} 
\def\physletb{{\rm Phys. Lett. B}}
\def\physrevc{{\rm Phys. Rev. C}} 
\def\physrevd{{\rm Phys. Rev. D}}
\def\sovphysjetp{{\rm Soviet~Phys.~JETP}} 
\def\ptpl{{\rm Progr.Theor.Phys.Lett}} 
\def\ptps{{\rm Prog.Theor.Phys.Suppl.}}
\def\ptp{{\rm Prog. Theor. Phys.}}

\begin{figure}
\caption{\label{fig:scales-fig} Scales relevant for the evolution of
Alfv\'en and slow magnetosonic waves before recombination, calculated
for modes that propagate at $\cos\theta=1$ and a background magnetic
field of $B_0 = 3\times 10^{-9} {\rm Gauss}$ today. All length scales
are given in comoving units. Any mode with fixed comoving wavelength
$\lambda_c$ will at cosmic temperature $T$ be in the photon diffusion
regime if it is to the left of the dotted line, or in the photon
free-streaming regime if it is to the right. Modes with wavelength
$\lambda_c$ will at temperature $T$ be non-oscillatory (overdamped) if
they are between the two solid lines. The two dashed lines indicate
the temperature at which a mode of given wavelength is damped by one
e-fold, either during its oscillatory evolution in photon diffusion or
its overdamped evolution in photon free-streaming. The figure assumes
$\Omega_b=0.0125$ and $h=1$, and equality between radiation and matter
energy density at $T_{\rm EQ}=5.5$eV. See \S 4.2 for a more detailed
explanation.}
\end{figure}

\begin{figure}
\caption{\label{fig:WKBfigure}The evolution of the Fourier amplitude
as a function of cosmic temperature calculated numerically (solid
line), and analytically using the WKB approximation (dashed line), for
Alfven waves with three different comoving wavelengths indicated on
the figure. For the calculation we assume $B_0 = 3\times 10^{-9} {\rm
Gauss}$, $\cos\theta=1$, $\Omega_b=0.0125$, and $h=1$, and we have
fixed the ionization fraction at $X_e = 1$ even for temperatures below
the approximate temperature of recombination $T\approx 0.25 eV$. The
largest scale which the numerical calculation shows to be damped by
one e-fold before recombination (at ), corresponds to the analytically
calculated length scale of $2\times 10^{23}\,{\rm cm}$.}
\end{figure}

\appendix
\section{Radiation Diffusion}
\label{sec:diff_app}
This appendix contains the diffusion regime dispersion relations and
their solutions. The dispersion relation for magnetosonic waves,
expanded to first order in $\kappa'$, $\eta'$, and $\xi'$ (which
corresponds to a first order expansion in $l_\gamma/\lambda$), is:

\begin{eqnarray}
&& \omega^5(6\kappa' (1+R) + 3 \kappa'\va^2)
\nonumber\\
&& {}+i\omega^4\bigl((1+R)^2 +\va^2 (1+R)\bigr) 
\nonumber\\
&& {}+\omega^3\Bigl(7\eta'(1+R)\biggl(\frac{k}{a}\biggr)^2
                    + 3\xi' (1+R)\biggl(\frac{k}{a}\biggr)^2
                    +\kappa'(R^2-2)\biggl(\frac{k}{a}\biggr)^2 
		    + 3\eta'\va^2 \biggl(\frac{k}{a}\biggr)^2 
		    + \kappa'\va^2(R-2)\biggl(\frac{k}{a}\biggr)^2 
                \nonumber\\
        \noalign{\hfill $\displaystyle 
        {} + \eta'\va^2\cos^2\theta\biggl(\frac{k}{a}\biggr)^2
	   + 3\xi'\va^2\cos^2\theta\biggl(\frac{k}{a}\biggr)^2 
	   - 2\kappa'\va^2\cos^2\theta\biggl(\frac{k}{a}\biggr)^2 
	   + 2\aoa R (1+R) + \aoa \va^2 R) \Bigr) $}
\nonumber\\
&& {}+i\omega^2\Bigl(- \frac{1}{3} (1+R)\biggl(\frac{k}{a}\biggr)^2
                     - \va^2 (1+R) \biggl(\frac{k}{a}\biggr)^2
                     - \frac{1}{3} \va^2\cos^2\theta
			\biggl(\frac{k}{a}\biggr)^2\Bigr) \nonumber\\
&& {}+\omega\Bigl(- \eta' \biggl(\frac{k}{a}\biggr)^4
                  - 3\eta'\va^2 \biggl(\frac{k}{a}\biggr)^4
                  - \kappa'\va^2 (1+R)\biggl(\frac{k}{a}\biggr)^4   
	          - \eta' \va^2\cos^2\theta\biggl(\frac{k}{a}\biggr)^4 
                                          \nonumber\\
        \noalign{\hfill $\displaystyle       
        {} - 3\xi' \va^2\cos^2\theta\biggl(\frac{k}{a}\biggr)^4
           + 2\kappa'\va^2\cos^2\theta\biggl(\frac{k}{a}\biggr)^4
           - \frac{1}{3}\aoa R \biggl(\frac{k}{a}\biggr)^2
	   - \aoa\va^2 R \biggl(\frac{k}{a}\biggr)^2 \Biggr)
                  $ }   \nonumber\\
&& {}+i \frac{\va^2\cos^2\theta}{3}\biggl(\frac{k}{a}\biggr)^4
	+ i R \kappa' \aoa \va^2 \biggl(\frac{k}{a}\biggr)^4
	+ 3 i \kappa' \eta' \va^2 \biggl(\frac{k}{a}\biggr)^6 = 0.
\end{eqnarray}
Since the scales of interest are smaller than the horizon scale, we
only keep $\dot{a}/a$ terms which are first order in $l_\gamma/d_{\rm
H}$ and $\lambda/d_{\rm H}$. In addition, note that there are two
terms in the last row which seem to be second order in the expansion
variables, although for some slow magnetosonic waves for which
$B_0\cos\theta$ is so small that $\va^2\cos^2\theta \ll B_0^2
\eta'\kappa'$ or $\va^2\cos^2\theta \ll B_0^2 \kappa' \dot{a}/a$,
these terms play a crucial role in determining the damping rates for
overdamped solutions.

We recover the dispersion relation for sound waves from the one for
magnetosonic waves, if $B_0 = 0$, or if the magnetosonic wave
propagates along the field lines, ${\bf k} \| {\bf B}_0$:
\begin{equation}
3\omega^3\kappa'+i\omega^2(1+R)+\omega\biggl[
4\biggl(\frac{k}{a}\biggr)^2\eta' +3\biggl(\frac{k}{a}\biggr)^2\xi'-
(1-R)\biggl(\frac{k}{a}\biggr)^2\kappa'+\aoa R \biggr]
-\frac{i}{3}\biggl(\frac{k}{a}\biggr)^2 = 0
\end{equation}

The dispersion relation for Alfv\'en waves is:
\begin{equation}
3\omega^3 \kappa'
+ i\omega^2\biggl(1+R+ \va^2\biggr)
+ \omega\biggl(\aoa R +3\eta'\biggl(\frac{k}{a}\biggr)^2\biggr)
- i\va^2\cos^2\theta\biggl(\frac{k}{a}\biggr)^2=0\ .
\end{equation}

The solutions to the dispersion relations are presented using the
following convention: the oscillatory part of the frequency is denoted by 
$\omega_0
= {\rm Re}\,\omega$, and the damping rate by $\omega_1 = {\rm
Im}\,\omega$. For Alfv\'en waves the solutions are:
\begin{eqnarray}
	\omega_0 &=& \pm \frac{1}{(1+R+\tilde{B}_0^2)}
                     \sqrt{\tilde{B}_0^2(1+R+\tilde{B}_0^2)
                     	\biggl(\frac{k}{a}\biggr)^2-\frac{1}{4}
                     	\biggl(\frac{\dot{a}}{a}\biggr)^2 R^2}\ \ ,
                     		\nonumber\\ 
	\omega_1 &=& \frac{1}{2\bigl(1+R+\va^2\bigr)}
                     \Biggl[3\eta'\biggl({k\o a}\biggr)^2 + 3\kappa'
                     \va^2\cos^2\theta\biggl(\frac{k}{a}\biggr)^2
                     \frac{1}{(1+R+\tilde{B}_0^2)} + \aoa R
                     \Biggr]\ .  	\label{eq:Alfven_sol}
\end{eqnarray} 
For magnetosonic waves, a general solution for a non-dissipative fluid
is:
\begin{eqnarray}
\omega_0 =\pm\frac{v_s}{\sqrt{2}}\biggl(\frac{k}{a}\biggr) & & 
           \Biggl[\frac{1+R + 3\va^2(1+R)
           +\va^2\cos^2\theta }{\bigl(1+R+\va^2\bigr)}
           \nonumber \\
    & &   {} \pm\biggl(\frac{\bigl(1+R+3\va^2(1+R)+\va^2\cos^2\theta
		\bigr)^2}{\bigl(1+R+\va^2\bigr)^2}
           -12\va^2\cos^2\theta\frac{\bigl(1+R \bigr)}
	   {\bigl(1+R+\va^2\bigr)}\biggr)^\frac{1}{2}\Biggr]^\frac{1}{2}. 
		\label{eq:18}
\end{eqnarray}
This solution contains two magnetosonic solutions: fast magnetosonic,
whose frequency is obtained by taking the plus sign, and slow
magnetosonic, using the minus sign. For weak magnetic fields
Eq.~\ref{eq:18} may be expanded to second order in $B_0$, and the
solutions become:
\begin{equation}
\omega_0^{\rm FM} = \pm\vs\biggl(\frac{k}{a}\biggr) 
		   	\biggl(1 + \frac{3 R + 2}{2 (1+R)} 
			\va^2\sin^2\theta\biggr)
                   + O(B_0^4)\ , 
\end{equation} 
and
\begin{equation}          
\omega_0^{\rm SM} = \pm v_A \biggl(\frac{k}{a}\biggr)
			\biggl(1-\frac{3 R + 2}{2(1+R)} 
		   	\va^2\sin^2\theta\biggr) 
		   + O(B_0^5)\ .
\end{equation}
The imaginary (dissipative) parts of the magnetosonic frequencies,
again to first order in ${\dot a}/a$, $\kappa'$, and $\eta'$, and to
second order in $B_0$ are:
\begin{eqnarray}
&&
\omega_1^{\rm FM}=
  \biggl(\frac{k}{a}\biggr)^2\biggl[\frac{2}{1+R}\eta'
              + \frac{3}{2(1+R)}\xi'
              + \frac{R^2}{2 (1+R)^2} \kappa'\biggr]
+ \frac{1}{2}\aoa\biggl(\frac{R}{1+R}\biggr)
\label{eq:mag_sonic1}
\end{eqnarray}
and 
\begin{eqnarray}
&& \omega_1^{\rm SM}= {} \frac{3}{2(1+R)} \biggl(\frac{k}{a}\biggr)^2
	\eta^{\prime} + \frac{1}{2}\aoa\frac{R}{1+R} 
 	+ \biggl(\frac{k}{a}\biggr)^2\biggl[\frac{3}{2}\kappa'\va^2
		\sin^2\theta 
	+\frac{3}{2}\kappa'\va^2\cos^2\theta\frac{R}{(1+R^2)^2}
\bigg] 
\label{eq:mag_sonic2} 
\end{eqnarray}
For sound waves Eq.~\ref{eq:mag_sonic1} reduces to:
\begin{equation}
\omega_0 = \frac{1}{\sqrt{3(1+R)}}\biggl(\frac{k}{a}\biggr) \ , \nonumber
\end{equation}
and
\begin{equation}
\omega_1 =
\biggl(\frac{k}{a}\biggr)^2\biggl(\frac{R^2}{2(1+R)^2}\kappa'
 + \frac{2}{1+R}\eta' + \frac{3}{2(1+R)}\xi' \biggr) 
 + \frac{1}{2} \aoa \frac{R}{(1+R)} \ ,\label{eq:sound}
\end{equation}
which reproduces the solution of Weinberg (1971) if the
expansion is neglected ($\dot{a}/a=0$ and $a=1$).

\section{Neutrino Free-Streaming}
\label{sec:nu_fs_app}
During neutrino free-streaming, the dispersion relations are: 
\begin{equation}
\omega^2(1+\va^2)-\frac{3}{4}i\alpha\omega-\va^2\cos^2\theta\biggl(\frac{k}{a}\biggr)^2=0
\end{equation}
for Alfv\'en waves, and 
\begin{eqnarray}
&&	\omega^4 (1+\va^2) 
	- i\omega^3 \frac{3}{4}\biggr[(2\alpha+\gamma)
		+(\alpha+\gamma)\va^2\biggr]
	\nonumber\\ 
&& {} 	- \omega^2 \biggl[\frac{1}{3}\biggl(\frac{k}{a}\biggr)^2 
		+ \va^2\biggl(\frac{k}{a}\biggr)^2 
		+ \frac{\va^2\cos^2\theta}{3}\biggl(\frac{k}{a}\biggr)^2
		+ \frac{9}{16}(\alpha^2+2\alpha\gamma
		+ \alpha\gamma \va^2) \biggr]
	\nonumber\\  
&& {}	+ i\omega \biggl[\frac{1}{4}\alpha\biggl(\frac{k}{a}\biggr)^2 
		+ \frac{3}{4}(\alpha+\gamma)\va^2\biggl(\frac{k}{a}\biggr)^2
		+ \biggl(\frac{3}{4}\biggr)^2\alpha^2\gamma\biggr]
	+\frac{\va^2\cos^2\theta}{3}\biggl(\frac{k}{a}\biggr)^2 
		+ \frac{9}{16}\alpha\gamma\va^2\biggl(\frac{k}{a}\biggr)^2 = 0
\end{eqnarray}
for magnetosonic waves.

The solution to the dispersion relation for oscillatory Alfv\'en waves
is:
\begin{equation}
	\omega^{\rm A}_{\rm osc} = \pm\frac{v_{\rm A}\cos\theta}
				   {\sqrt{1+\va^2}}\biggl(\frac{k}{a}\biggr)
				    + \frac{3}{8}i\frac{\alpha}{(1+\va^2)}.
\end{equation}

The oscillatory magnetosonic solutions are expanded to first order in
the dissipation coefficients $\alpha$ and $\gamma$, corresponding to
the first order expansion in $\lambda/l_\nu$. This yields for fast
magnetosonic waves:
\begin{equation}
\omega^{\rm FM}_{\rm osc} = \pm v_s (1+\va^2\sin^2\theta)
				\biggl(\frac{k}{a}\biggr)
		 	    +\frac{3}{8}i(\alpha+\gamma)\ ,
\end{equation}
and for slow magnetosonic waves:
\begin{equation}
\omega^{\rm SM}_{\rm osc} = \pm v_A \cos\theta
			(1-\va^2\sin^2\theta)\biggl(\frac{k}{a}\biggr)
		 +\frac{3}{8}i(\alpha
			+3\va^2\gamma )\ .
\end{equation}
The decay rates for overdamped slow manetosonic and Alfv\'en waves
are given in the text in Eq.~\ref{eq:1400} and Eq.~\ref{eq:1401}.
%

\section{Photon Free-Streaming}
\label{sec:ph_fs_app}
During photon free-streaming, the dispersion relation for Alfv\'en
waves is:
\begin{equation}
\omega^2-i\omega\biggl(\alpha+{\aoa}\biggr)-\frac{\tilde{B}_0^2 \cos^2\theta}
{R}\biggl(\frac{k}{a}\biggr)^2=0\ ,\label{eq:109}   
\end{equation} and the dispersion relation for magnetosonic waves is:
\begin{eqnarray}
&&\lefteqn{
\omega^5 - i\omega^4 \biggl( 2\alpha + \gamma +3\aoa \biggr)
- \omega^3 \biggl[c_s^2\biggl(\frac{k}{a}\biggr)^2
+ \frac{\tilde{B}_0^2}{R} \biggl(\frac{k}{a}\biggr)^2 
+ \alpha^2+2\alpha\gamma 
+ 2\aoa\bigl(2\alpha +\gamma\bigr)\biggl]} \nonumber\\
&& {} 
+ i\omega^2 \biggl[\alpha \biggl(c_s^2+\frac{\tilde{B}_0^2}{R}\biggr)
\biggl(\frac{k}{a}\biggr)^2
+ \gamma \biggl({3\over 5}c_s^2 + {\tilde{B}_0^2\over R}\biggr)
\biggl({k\over a}\biggr)^2
+ \aoa\biggl(c_s^2+2{\tilde{B}_0^2\over R}\biggr)\biggl({k\o a}\biggr)^2
+ \alpha^2\biggl(\gamma + \aoa\biggr)+2\alpha\gamma\aoa\biggr] \nonumber\\ 
&& {} 
+ \omega \biggl(\frac{k}{a}\biggr)^2 
\biggl[\frac{c_s^2 \tilde{B}_0^2\cos^2\theta}{R}\biggl(\frac{k}{a}\biggl)^2
+ \alpha\gamma \biggl(\frac{3}{5}c_s^2 + \frac{\tilde{B}_0^2}{R}\biggr)
+ \gamma\aoa \biggl(\frac{3}{5}c_s^2 + \frac{\tilde{B}_0^2}{R}\biggr)
+ \alpha\aoa \frac{\tilde{B}_0^2}{R} \biggr] 
- \frac{3}{5}i \gamma c_s^2\frac{\tilde{B}_0^2 \cos^2\theta}{R}
\biggl(\frac{k}{a}\biggr)^4 = 0.
\end{eqnarray}
The magnetosonic dispersion relation has been derived to first order in
$\dot{a}/a$, but without any approximation in the dissipative coefficients
$\alpha$ and $\gamma$.

The oscillatory solution to the dispersion relation for Alfv\'en waves is:
\begin{equation}
\omega^{\rm A}_{\rm osc} =\pm c_{\rm A} \cos\theta\biggl(\frac{k}{a}\biggr)
+{i\o 2}\biggl(\alpha+{\aoa}\biggr) \ .
\end{equation}
For fast magnetosonic waves in the adiabatic regime (${\rm
Re}\,\omega\gg\gamma ,\alpha$) the oscillatory solution is:
\begin{equation}
\begin{array}{rcll}
	 \omega^{\rm FM}_{\rm osc} & = & \displaystyle
		\pm c_s \biggl(1+{\tilde{B}_0^2\o 2c_s^2 R}\sin^2\theta\biggr)
		\biggl(\frac{k}{a}\biggr)
		+i\biggl({\alpha\o 2}+{\gamma\o 5}+{\aoa}\biggr) 
		&  \mbox{for $c_s \gg c_A$} \\ [0.2in]
	\omega^{\rm FM}_{\rm osc} & = & \displaystyle 
		\pm c_{A}\biggl(1+{c_s^2 R\o 2\tilde{B}_0^2}\sin^2\theta
		\biggr)\biggl(\frac{k}{a}\biggr)
		+i\biggl({\alpha\o 2}+{\gamma\o 5}{c_s^2Rsin^2\theta
		\o \tilde{B}_0^2} +{1\o 2}{\aoa}\biggr)
		& \mbox{for $c_s \ll c_A$}
\end{array}
\label{eq:113}
\end{equation}
and in the isothermal regime ($\gamma\gg {\rm Re}\omega\gg\alpha$) it is:
\begin{equation}
\begin{array}{rcll}
	 \omega^{\rm FM}_{\rm osc} & = & \displaystyle 
		\pm \sqrt{3\o 5}c_s \biggl(1+{5\o 6}{\tilde{B}_0^2\o 
		c_s^2 R}\sin^2\theta\biggr)\biggl(\frac{k}{a}\biggr)
		+i\biggl({\alpha\o 2}
		+{c_s^2\o 5\gamma}\biggl(\frac{k}{a}\biggr)^2
		+{1\o 2}{\aoa}\biggr)
		&  \mbox{for $c_s \gg c_A$} \\ [0.2in]
	\omega^{\rm FM}_{\rm osc} & = & \displaystyle 
		\pm c_{A}\biggl(1+{3\o 10}{c_s^2 R\o\tilde{B}_0^2}
		\sin^2\theta\biggr)\biggl(\frac{k}{a}\biggr)
		+i\biggl({\alpha\o 2}
		+{c_s^2\o 2\gamma}sin^2\theta\biggl(\frac{k}{a}\biggr)^2 
		+{\aoa}\biggr)
		& \mbox{for $c_s \ll c_A$}.
\end{array}
\label{eq:115}
\end{equation}
The condition for adiabaticity is dependent on the strength of the
magnetic field when $c_A\gg c_s$.

In the limit $B_0 = 0$, the above solutions give the solutions for
sound waves:
\begin{equation}
\omega = \pm c_s \biggl(\frac{k}{a}\biggr)+i\biggl({\alpha\o 2}+{\gamma\o 5}+{\aoa}\biggr)\
,\label{eq:110}
\end{equation}
in the adiabatic regime ($c_s k/a \gg\gamma ,\alpha$) and
\begin{equation}
\omega=\pm \sqrt{3\o 5}c_s \biggl(\frac{k}{a}\biggr)+i\biggl({\alpha\o 2}+
+{c_s^2\o 5\gamma}\biggl(\frac{k}{a}\biggr)^2+{1\o 2}{\aoa}\biggr)\ ,
		\label{eq:111}
\end{equation}
in the isothermal regime ($\gamma\gg c_s k/a \gg\alpha$).

The oscillation frequency and the damping rate for slow magnetosonic
waves in the adiabatic regime are:
\begin{equation}
\begin{array}{rcll}
	 \omega^{\rm SM}_{\rm osc} & = & \displaystyle 
		\pm c_{A}\cos\theta\biggl(1-{\tilde{B}_0^2\o 2Rc_s^2}
		\sin^2\theta\biggr)\biggl(\frac{k}{a}\biggr) 
		+i\biggl({\alpha\o 2}+{\gamma\o 5}
		{\tilde{B}_0^2\sin^2\theta\o c_s^2R} +{1\o 2}{\aoa}\biggr)
		&  \mbox{for $c_s \gg c_A$} \\ [0.2in]
	\omega^{\rm SM}_{\rm osc} & = & \displaystyle 
		\pm c_s \cos\theta\biggl(1-{c_s^2R\o 2\tilde{B}_0^2}
		\sin^2\theta\biggr)\biggl(\frac{k}{a}\biggr)
	+i\biggl({\alpha\o 2}
		+{\gamma\o 5}+{\aoa}\biggr)
		& \mbox{for $c_s \ll c_A$},
\end{array}
\end{equation}
and in the isothermal regime:
\begin{equation}
\begin{array}{rcll}
	 \omega^{\rm SM}_{\rm osc} & = & \displaystyle 
		\pm c_{A}\cos\theta\biggl(1-{5\o 6}{\tilde{B}_0^2\o 
		Rc_s^2}\sin^2\theta\biggr)\biggl(\frac{k}{a}\biggr) 
		+i\biggl({\alpha\o 2}
		+{1\o 2}{\aoa}\biggr)
		&  \mbox{for $c_s \gg c_A$} \\ [0.2in]
	\omega^{\rm SM}_{\rm osc} & = & \displaystyle 
		\pm \sqrt{3\o 5}c_s \cos\theta\biggl(1-{3\o 10}{c_s^2R\o 
		\tilde{B}_0^2}\sin^2\theta\biggr)\biggl(\frac{k}{a}\biggr)
		+i\biggl({\alpha\o 2}+{1\o 2}{\aoa}\biggr)
		& \mbox{for $c_s \ll c_A$},
\end{array}
\end{equation}
Note that all the solutions are derived for a
background magnetic field whose energy density is much smaller than
the energy density in photons ($\va\ll 1$).

In contrast to relativistic MHD there do exist non-relativistic, 
fast magnetosonic waves in the overdamped limit.
The decay rates for the amplitudes of these overdamped fast magnetosonic
waves are:
\begin{equation}
\begin{array}{rcll}
	 \omega^{\rm FM}_{od} & = & \displaystyle 
		i {3c_s^2\o 5\alpha}\biggl(\frac{k}{a}\biggr)^2
		&  \mbox{for $c_s \gg c_A$} \\ [0.2in]
	\omega^{\rm FM}_{od} & = & \displaystyle 
		i {c_A^2\o \alpha}\biggl(\frac{k}{a}\biggr)^2
		& \mbox{for $c_s \ll c_A$}.

\end{array}
\end{equation}
The decay rates of overdamped slow magnetosonic and Alfv\'en waves
are given in the text in Eq.~\ref{eq:1300} - Eq.~\ref{eq:130}.

\section{Damping Scales}
\label{sec:scales_app}
In this appendix we give the evolution of the damping scale as a
function of
temperature for Alfv\'en and slow magnetosonic modes in the early
universe. The temperature dependence of the damping scale of
fast magnetosonic waves is given in Section 4.1.
Before neutrino decoupling, the damping scale for
Alfv\'en and slow magnetosonic modes evolves as follows:
\begin{equation}
\lambda^{\rm A,SM} \simeq \left\{ \begin{array}{ll}
	 \displaystyle 2\times 10^{20} {\rm cm} \,   
              	\biggl({T\o {\rm MeV}}\biggr)^{-{5\o 2}}
              	\biggl({g_{r}\o 10.75}\biggr)^{-{3\o 4}}
              	\biggl({g_{\nu}\o 5.25}\biggr)^{1\o 2}
		\biggl({g_l+g_q\o 8.75}\biggr)^{-{1\o 2}}
		&  \mbox{for $T>T_1$} \\ [0.2in]
	\displaystyle 3 \times 10^{15} {\rm cm} \,   
		(B_9 \cos\theta)^{5\o 3}
              	\biggl({g_{r}\o 10.75}\biggr)^{-{1\o 3}}
              	\biggl({g_{\nu}\o 5.25}\biggr)^{-{1\o 3}}
		\biggl({g_l+g_q\o 8.75}\biggr)^{1\o 3}
		&  \mbox{for $T_1>T>T_2$} \\ [0.2in]
	\displaystyle 10^{18} {\rm cm} \,
		B_9 \cos\theta   
              	\biggl({T\o {\rm MeV}}\biggr)^{-{5\o 2}}
              	\biggl({g_{r}\o 10.75}\biggr)^{1\o 4}
              	\biggl({g_{\nu}\o 5.25}\biggr)^{-{1\o 2}}
		\biggl({g_l+g_q\o 8.75}\biggr)^{-{1\o 2}}
		&  \mbox{for $T_2>T>T_3$} \\ [0.2in]
	\displaystyle 	10^{17} {\rm cm} \,
		B_9 \cos\theta   
              	\biggl({g_{r}\o 10.75}\biggr)^{-1}
              	\biggl({g_{\nu}\o 5.25}\biggr)^{1\o 3}
		\biggl({g_l+g_q\o 8.75}\biggr)^{1\o 3}
		&  \mbox{for $T_3>T$}. \\ [0.2in]
\end{array}
\right.
	\label{eq:nu_templength}
\end{equation}

The damping scale at $T>T_1$ is approximately the same as the damping
scale for fast magnetosonic modes, and represents the diffusion length
of the decoupling particles. It is illustrated for photon decoupling
by the dashed line on the left-hand-side of
Fig.~\ref{fig:scales-fig}. The largest wavelength mode still damped by
one e-fold during oscillatory diffusion becomes overdamped at
temperature $T_1$:
\begin{equation}
	\biggl({T_1 \o {\rm MeV}}\biggr) = 80\,\,
		(B_9 \cos\theta)^{-{2\o 3}} 
		\biggl({g_{r}\o 10.75}\biggr)^{-{1\o 6}}
              	\biggl({g_{\nu}\o 5.25}\biggr)^{1\o 3}
		\biggl({g_l+g_q\o 8.75}\biggr)^{-{1\o 3}}\,,
\end{equation} 
i.e. the temperature at which the oscillatory damping scale 
(diffusion length) and the
overdamping length on the graph intersect.

Since the damping is inefficient during the overdamped diffusion
phase, the largest length scale damped in diffusion represents the
maximum damping scale, until further damping during
free-streaming. Temperature $T_2$ represents the point when the
overdamped free-streaming damping scale (equivalent to the dashed line
on the right-hand-side of Fig.~\ref{fig:scales-fig}) exceeds 
the maximum scale damped so far,
\begin{equation}
	\biggl({T_2\o {\rm MeV}}\biggr) = 10\,\,
		(B_9 \cos\theta)^{-{4\o 15}} 
		\biggl({g_{r}\o 10.75}\biggr)^{7\o 30}
              	\biggl({g_{\nu}\o 5.25}\biggr)^{-{1\o 15}}
		\biggl({g_l+g_q\o 8.75}\biggr)^{-{1\o 3}}\,.
\end{equation} 
If this happens before decoupling, the damping scale grows
further, and is now determined by the free-streaming damping
scale. Finally, the largest length scale damped in free-streaming is
the one which still damps by one e-fold before becoming oscillatory
again, determined by the intersection of the dashed and the solid line
on the right-hand-side of Fig.~\ref{fig:scales-fig}).
This happens at:
\begin{equation}
	\biggl({T_3\o {\rm MeV}}\biggr) = 5\, 
		\biggl({g_{r}\o 10.75}\biggr)^{1\o 2}
              	\biggl({g_{\nu}\o 5.25}\biggr)^{-{1\o 3}}
		\biggl({g_l+g_q\o 8.75}\biggr)^{-{1\o 3}}\,.
\end{equation} 
The final damping scale is determined by this transition; as it is
given in Eq.~\ref{eq:nu_templength} for $T<T_3$, it includes the
scales damped during the transition itself. Note that for large
magnetic field strength, with energy density approaching equipartition
with radiation energy density ($B_9 \approx 10^{3}$), and
$\cos\theta\approx 1$, the maximum damping length scale is determined
by damping during the oscillatory diffusion phase rather than during
the overdamped free-streaming phase. In this case the damping scale of
Alfv\'en and slow magnetosonic modes becomes similar to the damping
scale of fast magnetosonic modes.

Around photon decoupling, the damping scale for Alfv\'en modes, and
the damping scale for slow magnetosonic modes when $c_s>c_A$, have
the same form:
\begin{equation}
\lambda^{\rm A,SM} \simeq \left\{ \begin{array}{ll}
	\displaystyle 10^{26} {\rm cm} \, 
		\biggl(\frac{\Omega_b h^2}{0.0125}\biggr)^{-{1\o 2}} 
		\biggl({T\o 0.25{\rm eV}}\biggr)^{-{3\o 2}} 
		& \mbox{for $T>T_1$} \\ [0.2in] 
	\displaystyle 10^{21} {\rm cm} \, 
		(B_9 \cos\theta)^3
		\biggl(\frac{\Omega_b h^2}{0.0125}\biggr)
		& \mbox{for $T_2<T<T_1$} \\ [0.2in]\,
	\displaystyle 4 \times 10^{23} {\rm cm} \,
		B_9 \cos\theta
              	\biggl({T\o 0.25{\rm eV}}\biggr)^{-{3\o 2}}
              	\biggl(\frac{\Omega_b h^2}{0.0125}\biggr)^{-{1\o 2}}
		& \mbox{for $T_{\rm EQ}<T<T_2$}, \\ [0.2in]
	\displaystyle 2 \times 10^{23} {\rm cm} \,
		B_9 \cos\theta
              	\biggl({T\o 0.25{\rm eV}}\biggr)^{-{5\o 4}}
		\biggl(\Omega_0 h^2\biggr)^{-{1\o 4}}
              	\biggl(\frac{\Omega_b h^2}{0.0125}\biggr)^{-{1\o 2}}
		& \mbox{for $T<T_{\rm EQ}$}. \\ [0.2in]
\end{array}
\right.
	\label{eq:slow-first}
\end{equation}
On the other hand, if $c_A>c_s$, the damping scale for slow
magnetosonic modes evolves like:
\begin{equation}
\lambda^{\rm SM} \simeq \left\{ \begin{array}{ll}
	\displaystyle 10^{26} {\rm cm} \, 
		\biggl(\frac{\Omega_b h^2}{0.0125}\biggr)^{-{1\o 2}} 
		\biggl({T\o 0.25{\rm eV}}\biggr)^{-{3\o 2}} 
		& \mbox{for $T>T_1$} \\ [0.2in] 
	\displaystyle 10^{21} {\rm cm} \, 
		(B_9 \cos\theta)^3 
		\biggl(\frac{\Omega_b h^2}{0.0125}\biggr) 
		& \mbox{for $T_2'<T<T_1$} \\ [0.2in]\,
	\displaystyle 4 \times 10^{21} {\rm cm} \,
		\cos\theta  
              	\biggl({T\o 0.25{\rm eV}}\biggr)^{-{3\o 2}}
		& \mbox{for $T_{\rm EQ}<T<T_2'$}. \\ [0.2in]
	\displaystyle 3 \times 10^{21} {\rm cm} \,
		\cos\theta  
              	\biggl({T\o 0.25{\rm eV}}\biggr)^{-{5\o 4}}
		\biggl(\Omega_0 h^2\biggr)^{-{1\o 4}}
		& \mbox{for $T<T_{\rm EQ}$}. \\ [0.2in]
\end{array}
\right.
	\label{eq:slow-second}
\end{equation}
The transition temperatures are determined by matter-radiation
equality at $T_{\rm EQ}= 5.5\,{\rm eV} (\Omega_0 h^2)$, and by the
last scale damped by one e-fold in oscillatory diffusion, which becomes
overdamped at $T_1$,
\begin{equation}
	\biggl({T_1\o 0.25{\rm eV}}\biggr) = 2\times 10^{3} 
		(B_9\cos\theta)^{-2}  
		\biggl(\frac{\Omega_b h^2}{0.0125}\biggr)^{-1},
\end{equation}
and presents the largest damping scale until damping in free-streaming
damps even larger scales, at $T<T_2$ with
\begin{equation}
	\biggl({T_2\o 0.25{\rm eV}}\biggr) = 50 (B_9
		\cos\theta)^{-{4\o 3}} \biggl(\frac{\Omega_b
		h^2}{0.0125}\biggr)^{-1}\,,
\end{equation}
or $T<T_2'$ with 
\begin{equation}
	\biggl({T_2'\o 0.25 {\rm eV}}\biggr) = 2
		B_9^{-2} 
		(\cos\theta)^{-{4\o 3}} 
              	\biggl(\frac{\Omega_b h^2}{0.0125}\biggr)^{-{2\o 3}}\,.
\label{eq:1600}
\end{equation}
However, if $T_2<T_{\rm EQ}$ the damping scale of $10^{21} {\rm cm} \,
(B_9 \cos\theta)^3 (\Omega_b h^2/0.0125)$ is valid until $(T/0.25
{\rm eV}) = 2 B_9^{-12/5} \cos\theta^{-8/5} (\Omega_b
h^2/0.0125)^{-4/5} (\Omega_0 h^2/0.0125)^{-1/5}$ when it changes to $3
\times 10^{21} {\rm cm} \, \cos\theta (T/0.25{\rm eV})^{-5/4}
(\Omega_0 h^2)^{-1/4}$.


The above equations are derived for the case in which the universe is
radiation dominated during damping in oscillatory diffusion
($T_1>T_{\rm EQ}$). This is always true for the background magnetic
fields with $B_9 \leq 1$. Although for larger background magnetic
fields the damping processes are the same, the temperature dependence
of the diffusion damping scale and its time of overdamping might be
different, resulting in a different final damping scale. For example,
for $B_9 \cos\theta\approx 10^3$ (or $v_{\rm A} \cos\theta \approx
1$), the oscillatory diffusion damping scale dominates up to
recombination, and its evolution in the matter dominated era
determines the damping scale at recombination.

Slow magnetosonic modes which propagate almost perpendicular to the
background magnetic field have different damping scales due to the
additional term in Eq.~\ref{eq:slow_diff_overd}. However, this scale
is substantially different from those given in the text only for modes
with $\cos\theta <0.03 B_9^{8/3}$, which makes the influence of this
additional damping term on the overall damping of slow magnetosonic
modes negligible.

\end{document}